%% file: main_file_arxiv.tex
\documentclass[aps,pra,preprintnumbers,twoside,twocolumn,superscriptaddress,floatfix,10pt]{revtex4-2}

\usepackage{preamble}
\makeatletter
\newcommand*{\addFileDependency}[1]{
  \typeout{(#1)}
  \@addtofilelist{#1}
  \IfFileExists{#1}{}{\typeout{No file #1.}}
}
\makeatother

\newcommand*{\myexternaldocument}[1]{%
    \externaldocument{#1}%
    \addFileDependency{#1.tex}%
    \addFileDependency{#1.aux}%
}

\myexternaldocument{Sijournal}

\begin{document}

\title{Modeling the Influence of Solvation on the Electrochemical Double Layer of Salt / Solvent Mixtures}

\begin{abstract}
    Modelling electrolytes accurately on both a nanoscale and cell level can contribute to improving battery chemistries.\cite{ArmandNature2008} We previously presented a  thermodynamic continuum model for electrolytes.\cite{schammer2020theory} In this paper we include solvation interactions between the ions and solvent, which alter the structure of the electochemical double layer (EDL). We are able to combine a local solvation model -- permitting examination of the interplay between electric forces and the ion-solvent binding -- with a full electrolyte model. Using this, we can investigate double layer structures for a wide range of electrolytes, especially including highly concentrated solutions. We find that some of the parameters of our model significantly affect the solvent concentration at the electrode surface, and thereby the rate of solvent decomposition. Firstly, an increased salt concentration weakens the solvation shells, making it possible to strip the solvent in the EDL before the ions reach the surface. The strength of the ion-solvent interaction also affects at which potential difference the solvation shells removed. We are therefore able to qualitatively predict EDL structures for different electrolytes based on parameters like molecule size, solvent binding energy and salt concentration.
\end{abstract}

\author{Constantin Schwetlick}
\affiliation{Helmholtz Institute Ulm, Helmholtzstra{\ss}e 11, 89081 Ulm,
  Germany}  
  \affiliation{Universit\"at Ulm,
  Albert-Einstein-Allee 47, 89081 Ulm, Germany} 

\author{Max Schammer}
\affiliation{Helmholtz Institute Ulm, Helmholtzstra{\ss}e 11, 89081 Ulm,
  Germany}  
\affiliation{German Aerospace Center, Wilhelm-Runge-Straße 10, 89081 Ulm, Germany}

\author{Arnulf Latz}
\affiliation{Helmholtz Institute Ulm, Helmholtzstra{\ss}e 11, 89081 Ulm,
  Germany}  
  \affiliation{Universit\"at Ulm,
  Albert-Einstein-Allee 47, 89081 Ulm, Germany} 
\affiliation{German Aerospace Center, Wilhelm-Runge-Straße 10, 89081 Ulm, Germany}

\author{Birger Horstmann}
\email{birger.horstmann@dlr.de}
\affiliation{Helmholtz Institute Ulm, Helmholtzstra{\ss}e 11, 89081 Ulm,
  Germany}  
  \affiliation{Universit\"at Ulm,
  Albert-Einstein-Allee 47, 89081 Ulm, Germany} 
\affiliation{German Aerospace Center, Wilhelm-Runge-Straße 10, 89081 Ulm, Germany}

\maketitle 
\allowdisplaybreaks

\input{body_main_file}

\section*{Acknowledgment}
\label{sec:ackn}

This work was funded in the project POLiS by the Deutsche For\-schungsgemeinschaft (DFG) under Germany's Excellence Strategy – EXC 2154 – Project number 390874152.

\begin{figure*}[htb]
    \rule{\textwidth}{2pt}
    \centering
    \resizebox{\textwidth}{!}{\input{tikzfileGraphicalAbstract}}
    \caption{In battery operation, desolvating the cations before intercalation presents a kinetic energy barrier. Our continuum electrolyte model describes the solvation shell structure in the EDL as a function of the applied voltage between electrode and bulk electrolyte. In particular, it predicts a desolvation potential, above which the electric field is strong enough to strip the solvation shells in the EDL.}
    \label{fig:GraphicalAbstract}
    \rule{\textwidth}{2pt}
\end{figure*}

\bibliographystyle{apsrev4-2}
\bibliography{bibliography}	

\end{document}

%% file: body_main_file.tex

\section{Introduction}
\label{sec:intro}

Batteries play an important role in sustainable energy systems. It is necessary to optimise the electrolyte-electrode interface to improve battery operation.\cite{Edge2021Rev,XuElyteReview2014,ArmandNature2008} 

Predicting the structure of the ion solvation shell both in the bulk and in the EDL is of general interest. Among other effects, solvation has a profound influence on the electrolyte structure and behaviour. \cite{he2016, Cheng2022} In the electroneutral bulk, solvation effects affect ion transport. Furthermore, the structure of the EDL and the degree of coordination with the solvent near electrodes can have an important influence on the reaction kinetics occurring at the interface between the electrodes and electrolyte. For example at the electrode electrolyte interface the solvation shell must generally be removed for the ion to transfer into the electrode structure. On the other hand, the solvent can co-intercalate to lower the associated energy barrier compared to pure ion-intercalation. This can be seen in Sodium-battery systems with graphite / graphene electrodes and certain electrolytes,\cite{KimHaegyeom2015,Cohn2016} which was previously considered unsuitable due to graphite delamination, but recent research promises reversible operation,\cite{Goktas2018,Li2019}. Solvent co-intercalation can also be found in Zinc-batteries\cite{Liu2020,Borchers2021} -- in both systems, however, binding solvent in the electrode degrades ion-transport in the cell. As the Electrochemical Double Layer (EDL) is able to disturb the solvation shell, the interplay between the EDL-forces and solvation interaction determines the energy barrier needed to be overcome, when an ion is intercalated. Thus, this paper aims to more accurately model solvation structures in the EDL. 

Experimentally, it is possible to probe solvation shell structure via a number of methods. Raman and NMR spectroscopy can give insight into the solvent coordination number \cite{Cresce2017,Seo2015} Cyclic voltammetry (CV) can be used to extract the differential capacitance of an electrolyte at a charged surface and thereby probe the charge storage mechanisms in the EDL.\cite{Hamelin1983,Hamelin1985,Valette1989,Arulepp2004,Shatla2021} However, the concentration profiles cannot be measured at the scales of the EDL. Here, theoretical models can give a deeper understanding of the EDL-structure.

Due to the very small scale of the double layer, molecular dynamics (MD) simulation are used to investigate the structure. Vatamanu and Borodin have modelled water-in-salt electrolytes, where the EDL structure is strongly influenced by the applied voltage due to ion-solvent-interactions.\cite{Vatamanu2017}
Recently, continuum models have also been used to analyse EDL nanostructure. The groundwork was laid by Kornyshev, when he illustrated how to predict differential EDL capacitances using a modified Poisson-Boltzmann model.\cite{Kornyshev2007} 
Recently, the authors have published a framework based on consistent non-equilibrium thermodynamics.\cite{Hoffmann2018,schammer2020theory} 
Our model has been applied to different systems, including highly concentrated electrolytes used in batteries\cite{schammer2020theory,kilchert2024silicon} and electrolyte mixtures composed of a pure ionic liquids (ILs) mixed with a Li salt.\cite{doi:10.1021/acs.jpclett.2c02398,D2CP04423D}  
However, to investigate the nanostructure of the electrochemical double layer (EDL) -- like charge layering -- the model was supplemented by non-local interactions, based on a potential in style of classical density-functional theory. This extended model describes the formation of equilibrium structures in the EDL of ILs near electrified interfaces, and  replicates results from MD-simulations, including oscillations in charge density (``overscreening'').\cite{Hoffmann2018,schammer2021role} 
De Souza from the Bazant group introduced a Poisson-Boltzmann type model informed by similar non-local, hard-shell interactions.\cite{Souza2020}

There have been efforts to include solvation in continuum models. Dreyer and his Coworkers were able to study the effect of including ion-solvent interactions on key transport parameters.\cite{dreyer2019,DREYER201475}  The stabilising effects of the solvent in confined geometries has been studied in aqueous electrolytes using MD.\cite{Fong2024} For a more detailed description of the solvation shell structure, DFT models can be used.\cite{Cresce2017} This allows estimation of the binding energy and preferred coordination numbers, key parameters necessary for the continuum scale models. Based on these parameters, Goodwin developed a mean field model, able to describe complex clusters containing both ions and solvent molecules. \cite{Goodwin2023} Their detailed work predicts the association probabilities of two electrolyte species, from which the composition of solvation clusters can be inferred. 

In this manuscript, we present a theory for the solvation of ions by a neutral solvent. Our continuum model presented here extends a transport theory published earlier by some of the authors.\cite{schammer2020theory} The underlying framework is based on modelling the free energy of the system. Here, we modify the free energy and include solvation effects, via accounting for the solvation binding energy and entropic effects. Thereby, our model captures the interplay between the strong electric forced in the EDL and the binding forces in the solvation shells. From this, we can investigate the stripping of the solvation shell around the ions, which is prompted by strong electric fields. We can give qualitative insight, which electrolyte properties can improve intercalation kinetics and reduce solvent decomposition at the electrode. 

We structure this manuscript as follows. First, in \cref{sec:model} we describe our thermodynamic transport model based on a modified free energy which includes solvation. In \cref{sec:comp} we describe the numerical implementation used to solve the differential equations. We use the results from simulations of the EDL structure to compare the model with and without solvation in \cref{sec:results_difcap}. Next, in \cref{sec:discussion} we present a parameter study where we investigate the effect of the ion-solvent binding energy as well as size asymmetry between cat- and anions on the EDL structure. Additionally, we compare our results to experimental differential capacitances obtained from CV. A summary of our work is found in \cref{sec:conclusion}.

\section{Model}
\label{sec:model}

In this chapter, we present our continuum electrolyte model for the solvation of ions by neutral solvent molecules. This chapter has three parts: First, in \cref{sec:transporttheory}, we discuss our continuum description based on modeling the free energy for a general electrolyte composition. Second, \cref{sec:three_comp_system} applies our model to a three species system of a binary salt with a neutral solvent.

\subsection{Electrolyte Transport Theory including solvation}
\label{sec:transporttheory}

Our theory is based on a transport theory for liquid electrolytes, which has recently been developed by the authors.\cite{Latz2011,schammer2020theory,schammer2021role} This framework combines elements from non-equilibrium thermodynamics, mechanics and electromagnetic theory, and focuses on modeling the free energy density of the system. Here, we incorporate solvation effects into our framework by adding additional terms to the bulk free energy discussed in Ref.~\citenum{schammer2020theory}. 

Because we neglect non-local interactions in our discussion, the Helm\-holtz free energy 
of the system, $F=\int \ce{d}V \massdens\freedens$, is a function of 
the free energy density $\massdens\freedens $.\cite{schammer2020theory} 

The quantity $\massdens\freedens $ constitutes the focal quantity in our modeling approach, as it accounts for the material specific properties of the system. This approach has the advantage, that the framework can easily be modified and adapted to different systems via modification of the free energy density. As a consequence, the framework exhibits a great potential for tunability. In this work, our model for the free energy density relies on the definition of
$\massdens\freedens$ in Ref.~\citenum{schammer2020theory}. 

We allow solvation of ions by neutral solvent molecules. For this purpose, we assume complete dissociation of the electrolyte salts into pure ionic species consisting of cations and anions. The neutral solvent species then partially solvates both the cations and the anions. 

The species concentrations $\conc_\alpha$ characterise the local electrolyte composition. In this manuscript, we denote the total concentration of the solvent species, which comprises all solvent molecules (bound and unbound), by $\conc_{\solvent}$. The specific volume fractions of all electrolyte species sum up to unity,\cite{schammer2020theory}  
\begin{equation}
    \label{eq:euler}
    \sum_{\alpha=1}^{\ce{N}} \pmv_\alpha \conc_\alpha 
    = \pmv_{\solvent}\conc_{\solvent} + \sum_{\alpha\neq \solvent}^{\ce{N}} \pmv_\alpha \conc_\alpha=1,
\end{equation}
where $\pmv_\alpha$ are the partial molar volumes of the species. Thus, only N-1 species concentrations in an  electrolyte mixture composed of N constituents are independent. 

Our description for the solvation of the cations and of the anions assumes that 
each cation and anion is coordinated by $\cordnumb_\alpha$ solvent molecules (where $\alpha\neq \ce{s}$). 
Thus, the concentration of ``free'' solvent (not bound to the ions), reads
\begin{equation}
    \label{eq:def_coordnumb}
    \concfree_{\solvent}=\conc_{\solvent} - \sum_{\alpha\neq \ce{s}}^{\ce{N}} 
\cordnumb_\alpha \conc_\alpha. 
\end{equation}
The solvent coordination numbers $\cordnumb_\alpha$  depend on the configuration of the system and are functions of the system variables. For simplicity, we assume discrete solvation shells with a fixed number of degenerate solvation-sites $\cordnumb_\alpha^{\ce{m}}$. The parameters $\cordnumb_\alpha^{\ce{m}}$ constitute the fixed upper bounds for the normalized coordination numbers $\tilde{\cordnumb}_\alpha = \cordnumb_\alpha/\maxcoord_\alpha$ (such that $\num{0}\leq\tilde{\cordnumb}_\alpha\leq\num{1}$. Thus, $\cordnumb_\alpha\to\num{0}$ corresponds to empty solvation shells, and $\cordnumb_\alpha\to\num{1}$ corresponds to solvation shells which are completely filled. Furthermore, we comprise all ``free'' molecules (ions and solvent) in
\begin{equation}
    \concfree=\concfree_{\solvent} +\sum_{\alpha\neq \solvent}^{\ce{N}} \conc_\alpha .
\end{equation}

The formation of equilibrium structures in the EDL is governed by the competition between disordering effects of entropy with ordering effects, e.g. due to electrostatics.\cite{schammer2021role} As consequence, the relative magnitudes of the corresponding energy scales determine the EDL. For non-interacting electrolytes, the relevant energies appearing in $\massdens\freedens$ are the Coulomb energy and the mixing entropy $\massdens\freedens^{\ce{mixing}}$. However, solvation effects lead to additional energy contributions. 

First, the solvation of ions implies a contribution to $\massdens\freedens$ via the specific ion-solvent binding energies. We model this solvation energy by $\massdens\freedens^{\ce{solvation}} = \sum_\alpha E_\alpha  \cordnumb_\alpha\conc_\alpha$, where the specific binding energies $E_\alpha$ are fixed parameters. In principle, they can be obtained from atomistic modeling (\textit{e.g.}, DFT or MD).\cite{Okoshi2013,Cui2016,Liu2020Li,Liu2020Na,Seo2012,Seo2015} 

In addition, the solvation of ions by neutral solvent molecules alters the statistics of the system and thus has an influence on the mixing entropy $\massdens\freedens^{\ce{mixing}}$, which depends on the number of microscopic realisations of a given macroscopic state. Usually, for a bulk electrolyte, it is modelled in analogy to an 
ideal gas.\cite{schammer2020theory} Because this approach is insufficient for our case, we derive a more accurate model based on modified statistics. 

In particular, solvation processes influence the mixing entropy $s^{\ce{mixing}}$,\cite{DREYER201475} which affects the free energy according to the constitutive equation $s^{\ce{mixing}}= -\p\freedens/\p T$ (here we neglect thermal contribution to the free energy).\cite{schammer2020theory} We model the mixing entropy 
by assuming that it obeys Boltzmann's law, i.e. $\massdens s^{\ce{mixing}}= R\ln(\Omega)/V$. Here, $\Omega$ is the number of possible microstate configurations for a given macrostate, all of which are assumed to be equally likely.  Furthermore, we assume that any given macrostate is unique\-ly defined by a fixed set of values for the species concentrations $\conc_\alpha$ and coordination numbers $\cordnumb_\alpha$.  The total number of microstates, however, is determined by the statistics of the free and bound particles via $\Omega = \Omega_{\ce{free}}\cdot \Omega_{\ce{bound}}$. For the free solvent molecules and ions this reads
\begin{equation}
\label{eq:omega_free}
    \Omega_{\ce{free}} = \frac{N^{\ce{free}}!}{ 
    N_{\solvent}^{\ce{free}}! \cdot  \prod_{\alpha\neq \solvent}(N_\alpha!)}.
\end{equation}
The total number of available sites for the solvent molecules in the ion solvation shells is for each ion species $\maxcoord_\alpha N_\alpha$. However, out of this number, only $\cordnumb_\alpha N_\alpha$ sites are filled (note that for the specific macrostate we only specify the average ion-solvent coordination number and do not resolve the distribution of the bound solvent molecules between the solvation shells).  Thus,
\begin{equation}
\label{eq:omega_bound}
    \Omega_{\ce{bound}} = \prod_{\alpha\neq\solvent} \frac{\left(\maxcoord_\alpha N_\alpha\right)!}{\left(\cordnumb_\alpha N_\alpha\right)!\cdot \left(\left[\maxcoord_\alpha-\cordnumb_\alpha\right]N_\alpha\right)!}.
\end{equation}
The mixing entropy can then be calculated directly from $\ln (\Omega)$ using Stirling's approximation,  $\ln N_\alpha! \approx N_\alpha \ln N_\alpha - N_\alpha$.

Altogether, the free energy including solvation effects reads (where $\nondim{E}_\alpha=E_\alpha/RT$)
\begin{multline}
\label{eq:bulk_freeenergy_all}
    \massdens \freedens =
     RT \sum_{\alpha\neq \solvent}^{\ce{N}} \conc_\alpha \maxcoord_\alpha
    \left(
        \tilde{\cordnumb}_\alpha \ln(\tilde{\cordnumb}_\alpha ) 
        + (1-\tilde{\cordnumb}_\alpha) \ln(1-\tilde{\cordnumb}_\alpha)
        + \tilde{\cordnumb}_\alpha\tilde{E}_\alpha
       \right)
       \\
       + RT \left( 
         \sum_{\alpha\neq \solvent}^{\ce{N}} \conc_\alpha \ln\frac{\conc_{\alpha}}{\concfree} 
         + \concfree_{\solvent}  \ln\frac{\concfree_{\solvent}}{\concfree}
    \right)
         \\
       + \frac{\bolds{ED}}{2} 
         +  \frac{\mathscr{K}}{2} \left( 1- \sum_{\alpha=1}^{\ce{N}}\conc_\alpha\pmv^0_\alpha \right)^2.
\end{multline}
The contributions appearing in the first and second line above comprise the mixing entropy and the binding energy. The solvation shells of the ions are filled with neutral solvent molecules such that the free energy becomes minimal. Thus, the fractional coordination numbers $\tilde{\cordnumb}_\alpha$ are governed by the constraint that $\partial(\massdens\freedens)/\partial\tilde{\cordnumb}_\alpha =0$. In the SI (see \cref*{sec:solution_solvation}), we show that this condition is fulfilled exactly if  
\begin{gather}
    \label{eq:extremal_condition_coordnumb}
    \tilde{\cordnumb}_\alpha 
    = \frac{\xparam}{\xparam +\exp\left(\tilde{E}_\alpha\right)},
    \intertext{with the molar ratio }
    \label{eq:def_xparam}
    \xparam = \concfree_{\solvent}/\concfree.
\end{gather}
The third term in \cref{eq:bulk_freeenergy_all} describes the electrostatic energy density of a linear dielectric medium, where $\bolds{D} = \diel\dielrel \bolds{E}$.\cite{schammer2020theory}
The final term measures the volumetric energy density of deformations, where  
the bulk-modulus $\mathscr{K}$ acts as a Lagrange multiplier in 
the case of incompressible electrolytes.\cite{schammer2020theory} In that case, the partial molar volumes are constant and equal the reference bullk values $\pmv_\alpha^0$. This term results in a pressure force, which constitutes a threshold for local ion concentrations and stabilises the electrolyte against attractive electrostatic forces.\cite{schammer2021role}

\subsection{Electrolyte Mixtures: Dilution of a salt by a neutral solvent}
\label{sec:three_comp_system}

In this section, we apply the theory derived above and focus on electrolyte mixtures composed of a completely dissociated salt with a neutral solvent species ranging from infinitely dilute systems (only solvent) up to pure salts (no solvent, e.g. ionic liquids). 

Here, we assign the cations and anions to the first and second species, i.e. $\conc_1=\conc_+$ and $\conc_2=\conc_-$, and choose for the third species the neutral solvent, i.e. $\conc_3 = \conc_\solvent$. As before, the volume fractions 
$\conc_\alpha\pmv_\alpha$ of the three species sum to 1 (\cref{eq:euler}).\cite{schammer2020theory} 
In contrast to the specific partial molar volumes $\pmv_\alpha$ (which are assumed constant in the incompressible limit), the partial molar volume of the solvated ions, given by $\pmv_\alpha+\cordnumb_\alpha
\pmv_{\solventlabel}$, depend on position (even for incompressible electrolyte 
mixtures). 

Volumetric effects often depend only on the size asymmetry between the species. Here, we use the partial molar volume of the salt $\pmvil = \pmv_\cationlabel + \pmv_\anionlabel$ as reference and describe the size asymmetry by ratios 
$\pmvrel_\alpha=\pmv_\alpha/\pmvil$. Note that, by construction, 
$\pmvrel_++\pmvrel_-=1$. 
Thus, for symmetric ions where $\pmvrel_+=\pmvrel_-$, the solvent is much smaller (larger) than the ions if $\pmvrel_\solvent\gg 1$ ($\pmvrel_\solvent\ll 1$), and has a volume similar to the salt if $\pmvrel_\solvent\approx 1$.

Because of the Euler equation for the volumes (see \cref{eq:euler}) only two species concentrations are independent. In addition, we replace one species concentration by the charge density $\chargedens= z_+ F(\conc_+-\conc_-)$. Thus, $\chargedens$ and $\conc_{\solventlabel}$ are the independent variables for the electrolyte configuration, where $ \conc_\pm(\conc_{\solventlabel},\chargedens)  = 1/\pmvil \pm \pmvrel_\mp\chargedens/Fz_+
- \pmvrel_\solvent\conc_{\solventlabel}$.\cite{schammer2020theory}

The characterization of an electrolyte mixture according to the amount of solvent depends on the initial composition. In solvent free electrolytes (e.g., ionic liquids), the initial configuration $\concbulk_\pm = \concbulk_{\ce{salt}}$ is governed by the condition of electroneutrality and by \cref{eq:euler}, such that the initial volume-fraction of the salt is $ \dilution=1$, where 
\begin{equation}
\label{eq:dilution}
 \dilution = \concbulk_{\ce{salt}}\pmvil.    
\end{equation}
Hence, the initial conditions of the electrolyte are fixed by the partial molar volumes of the ions. In contrast, when a solvent is present in the electrolyte, the initial volume-fraction of the salt is reduced according to \cref{eq:euler}
\begin{gather}
\label{eq:solvent_bulkconc}
    1 = \dilution + \concbulk_{\solventlabel}\pmv_{\solventlabel},
\end{gather} 
such that $\dilution <\num{1}$. As consequence one initial bulk concentration must be specified as additional parameter. 
By construction, $0\leq \dilution \leq 1$, where $\dilution\to 0$ corresponds to a pure solvent, 
and $\dilution\to 1$ corresponds to the solvent-free limit of a pure salt. 
Thus, the quantity $\dilution$ is a measure for the ``dilution'' of the electrolyte. 

In an electrolyte consisting of a solvent with a binary salt, the forces governing the particle fluxes are given by (see SI \cref*{sec:derivation_forces} for a derivation) 
\begin{align}
\label{eq:force_dimensional}
    \force_{\alpha\neq \solvent} = 
    z_\alpha F\bn\elpot 
    &+ RT 
    \bn \Biggl(
    \ln(1-\xparam) 
    - \maxcoord_\alpha \ln\left(\xparam+\exp[\tilde{E}_\alpha ]\right) \nonumber\\
    &\qquad\qquad
    - \frac{\pmvrel_\alpha}{\pmvrel_{\solvent}} \ln(\xparam)
    + \ln\frac{\conc_\alpha}{\conc_{\ce{IL}}}
    \Biggr),
\end{align}
with $\conc_{\ce{IL}} = \conc_\cationlabel + \conc_\anionlabel$. In the SI, we present a formulation of our model where all physical dimensions have been removed (see \cref*{sec:nondim}).

\section{Numerical Methods and Parameterisation}
\label{sec:comp}

\begin{figure}[tb]
    \centering
    \resizebox{\columnwidth}{!}{\input{tikzfileNumericalScheme}}
    \caption{This scheme illustrates our one-dimensional numerical set-up. For our simulations of the EDL structures, we assume the stationary state, in which the charge flux $\Jflux$ and particle flux $\Nflux_\alpha$ both vanish. Here, we consider a half-cell set-up, where the distance from the electrode (located at $x=\num{0}$) is measured along the x-axis. The right side of the simulation space is determined by the electroneutral bulk conditions, where $\chargedens^{\ce{bulk}} = 0$ and $\conc_{\solventlabel}=(1-\dilution)/\pmv_\solventlabel$. The electrostatic forces are governed by the applied voltage $V^\textsf{ext}$, given as the difference between the electrode potential and the bulk potential.}
    \label{fig:NumericalScheme}
\end{figure}

In this section, we describe our numerical implementation and pre\-sent our parameterisation. 

We solve the transport equations in a stationary state. \Cref{fig:NumericalScheme} illustrates the simulation space and boundary conditions used. The electrode located at $x=\num{0}$ is assumed to be completely inert, so no particles and therefore fluxes are allowed to cross the electrolyte-electrode interface. The other boundary is set by the contact with the bulk electrolyte at $x=x^{\ce{bulk}}$, which is electroneutral and at a potential of \SI{0}{\volt}, but able to exchange particles with the system. The ion concentrations on the bulk side are determined by a neutral charge density $\chargedens^{\ce{bulk}} = 0$ and the salt concentration $c_{\ce{salt}}^{\ce{bulk}} \pmv_{\ce{salt}} = \dilution$. The one-dimensional simulation space is spanned between these two boundaries, closed on the electrode side, open on the bulk side. 

The equilibrium structures in the EDL enforced by the stationary state $\Jflux, \Nflux = \boldsymbol 0$ require vanishing forces,\cite{schammer2021role} i.e.  $\force_\pm=0$, 
 stated in \cref{eq:force_dimensional}. As can be inferred from \cref{eq:force_dimensional,eq:def_xparam}, the forces are linear functions of gradients of the free variables,  $
    (\force_\cationlabel, 
    \force_\anionlabel)
    \propto 
        (\bn\elpot, 
        \bn\chargedens , 
        \bn\conc_{\solventlabel})$.
In contrast to the forces, which involve only first order derivatives, the Poisson equation $\chargedens=-\diel\bn( \dielrel \bn \elpot)$ involves a second order derivative of the potential. We linearise our system of equations and split the Poisson equation into two differential equations of first order,
\begin{equation}
    \Efield = -\bn\elpot, \quad \textnormal{and} \quad 
    \chargedens = \diel\bn\left(\dielrel\Efield\right).
\end{equation}
Altogether, the corresponding linearised fundamental system reads 
\begin{equation}
\label{eq:numerical_setup}
    \begin{pmatrix}
        \force_\cationlabel ,
        \force_\anionlabel,
        \Efield,
        \chargedens
    \end{pmatrix}^{T} = A( \chargedens,\conc_{\solventlabel}) \cdot 
    \begin{pmatrix}
    \bn\elpot ,
    \bn\chargedens,
    \bn\conc_{\solventlabel},
    \diel\bn\dielrel\Efield
    \end{pmatrix}^{T},
\end{equation} 
where the matrix $A$ is a function of the free variables $\{\elpot,\chargedens,\conc_{\solventlabel}\}$, but not of their gradients (stated in the SI \cref*{sec:num_implementation}). Note that, for the equilibrium state of the system, this fundamental system is subject to the condition $\force_{\pm}=0$.  

We solve the fundamental system of linear equations given by \cref{eq:numerical_setup} to obtain the local gradients as a function of the free variables, $ (\bn\elpot ,    \bn\chargedens, \bn\conc_{\solventlabel}, \diel\bn\dielrel\Efield)^{\ce{T}}=A^{-1} \cdot (0,0,
   \Efield ,
\chargedens)^{\ce{T}}$. This determines the value of the free variables at all points, if one configuration is known. The known configuration is given in the bulk by the electrolyte composition. However, to obtain a nontrivial solution, a small non-zero electric field has to be applied on the boundary. 

By calculating the local derivatives from the bulk configuration, we obtain the static solution for the whole double layer. The complete simulation data is then cut off at the predetermined, externally applied potential difference between bulk electrolyte and electrode $V^{\ce{ext}}$. Using this approach, the solution for any external voltage applied to the specific electrolyte can be obtained from one simulation.


According to this approach, the variables act as field quantities which depend on the position $x$ (distance from the electrode). However, there exists a local correspondence between the position $x$ and the electric potential within the electrolyte $\elpot(x)$. Thus, the position can be represented by the potential, and the EDL structure can therefore equally be described as a function of potential rather than space. This interpretation has the advantage that physical system parameters are removed from the functional description of the EDL properties. For example, as will be shown later, some structural properties (like the point at which the EDL becomes saturated with the counter-ion) occur at similar potentials over a wide range of electrolytes. Consequently, comparisons between different simulations are best done using the local potential.

Note that, according to this perspective, $\elpot=\SI{0}{\volt}$ corresponds to the bulk position $x^{\ce{bulk}}$. 

In the following sections, if not stated otherwise, we parameterise our electrolyte as follows. We assume symmetric ion sizes, by setting  $\pmvrel_\pm = \pmvrel_\solventlabel = 0.5$, and set the partial molar volume of the salt to $\pmvil = \SI{5e-5}{\meter^3\per\mol}$. Furthermore, we assume for the dielectric parameter $\dielrel = 50$, motivated by a DMSO solvent.\cite{Shatla2021} We set the number of available sites in the solvation shells to $\maxcoord_\pm=\num{4}$, which is commonly seen for many electrolytes. \cite{Liu2020Li,Liu2020Na} Although parameters for the binding energy can be obtained from atomistic models, the literature data is limited to a few systems which have been studied. Here, we set $\tilde{E}_\pm= -\num{4}$, inspired by DFT calculations for Sodium ions in carbonate solvents. \cite{Okoshi2013}

\section{Simulation results for the differential capacitance of the EDL}
\label{sec:results_difcap}

In this section, we present simulations results of the as-modeled solvation effects occurring in an electrolyte mixture composed of a salt and a neutral solvent. 

In our discussion of the numerical results, we focus mainly on the differential capacitance (DC) $\diffcap$ of the EDL. This quantity is a widely-used descriptor for the EDL structure, and has the advantage that it can be investigated both by experiments,\cite{Shatla2021} and by computer simulations.\cite{Kornyshev2007,Fedorov2014,Goodwin2020,Oldham2007} Here, we define it via $\diffcap = \partial \mathscr{Q}^{\ce{EDL}}/\partial V^{\ce{ext}}$, where $\mathscr{Q}^{\ce{EDL}}=\int_0^{x^{\ce{bulk}}} \chargedens(y) \ce{d}y  $ is the charge stored in the EDL, and $V^{\ce{ext}}$ is the external electrode potential. 

We structure this part into three sections. First, in \cref{sec:results_intro} we present a first glance on the numerical results for the primary variables of our model (electric potential and species concentrations). In the next sections of this part, we make use of these variables and calculate the corresponding differential capacitance. Second, in \cref{sec:analysis_nosolvation}, we aim to understand the influence of electrolyte dilution on the EDL structure. To address this goal, we present simulation results for our model where we neglect solvation effects by setting $\maxcoord_\pm=0$. In \cref{sec:analysis_solvation}, we then perform numerical simulation of our complete model taking solvation effects into account ($\maxcoord_\pm\neq 0$).  

\subsection{First Glance: Structure of the electrochemical double layer}
\label{sec:results_intro}

\begin{figure*}[tb]
    \begin{center}
    \labelfig[width=\columnwidth]{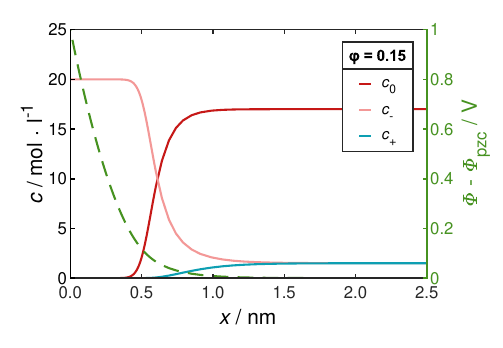}{fig:profile_nosolv}
    \labelfig[width=\columnwidth]{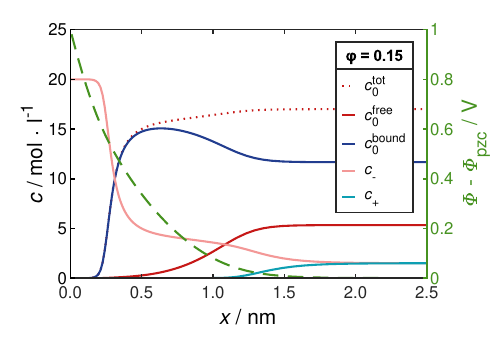}{fig:profile_solv}
    \caption{Concentration profiles (shown on the left y-axis), and profiles of the electric potential (dashed curves shown on the right y-axis) arising in the EDL at an inert electrode with an applied voltage of $V^{\ce{ext}}=\SI{+1}{\volt}$. \Cref{fig:profile_nosolv} shows the EDL structure when not including solvation interactions, while \cref{fig:profile_solv} shows numerical results when solvation effects are taken into account. Both figures show the concentration profiles for the solvent, and the two ion species, where the solvent is split into free and bound molecules when accounting for sol\-vation.     
    }
    \label{fig:example_profiles} 
    \end{center}
\end{figure*}

The simulations output the concentration profiles for the three electrolyte species (solvent, cation and anion), the local electric potential and the electric field as functions of the distance from the electrode. In this section, we aim to familiarize the reader with the properties of the system and highlight some typical findings about the EDL nano-structure.

\Cref{fig:example_profiles} shows numerical results for the concentration profiles of the electrolyte species (left y-axis), and of the electric potential (right y-axis) at an applied voltage of \SI{1}{\volt} for a moderately dilute electrolyte ($\dilution=\num{0.15})$. In the left figure (\cref{fig:profile_nosolv}), all solvation effects are neglected by setting $\maxcoord_\pm=\num{0}$, whereas the right figure (\cref{fig:profile_solv}) shows results including solvation effects (here, $\maxcoord_\pm =\num{4}$). Apparently, in both cases, the positively charged electrode is screened in the electrolyte by the accumulation of counter-ions (here, the anions), while repelling the cations. The unbound neutral solvent is also forced away from the electrode surface, albeit not as harshly as the cations. When solvation effects are neglected, the solvent does not get significantly closer to the electrode than the cations (see \cref{fig:profile_nosolv}). However, when solvation interactions are included (see \cref{fig:profile_solv}),  the behaviour of the solvent changes significantly, as solvent molecules bound to the solvation shell of the anions are now pulled much further into the EDL (closer to the electrode). The reaction rate of the solvent at the electrode surface can therefore be fine-tuned by modifying the solvation interaction.

For both cases presented in \cref{fig:example_profiles} (with and without solvation), the concentration profiles exhibit two to three distinct electrolyte regions with respect to the EDL structure. The first region is adjacent to the electrode, and is characterized by a saturation layer of the counter-ion accompanied by an almost complete depletion of the other ion species. Note that, because of the volume constraint in \cref{eq:euler}, the anion concentration is limited by the specific partial molar volume via $\conc_{-}^{\ce{sat}}=1/\pmv_{-}= \SI{20}{\mol\per\liter}$.\cite{schammer2021role} A second region can be identified on the opposite side far away from the electrode (in the electroneutral region of the bulk electrolyte), and is characterized by constant species concentrations. When the electrode is approached from this bulk region, however, a diffuse layer appears, in which the concentrations of the ions grow / decay exponentially. Meanwhile, the solvent concentration remains more or less constant. In this intermediate region, a quasi-saturation layer forms, where anions with completely filled solvation shells are in saturation (note that the effective size of the solvated ion is larger than the size of the unsolvated ions, which reduces the saturation concentration of the ions - see also \cref{eq:effective_volume}). 

For each of these regions a different energy scale shapes the local structure. \cite{schammer2021role} In the saturated region near the electrode, the electrostatic energy scale dominates both the entropic energy scale and the interaction energy scale. The diffusive structures emerging in the second region suggest that entropic effects are dominant in the bulk regime. A third region located between the first two regions emerges in the case where solvation effects are considered. Here, the bound solvent is the most prevailing species, and there are only small amounts of free solvent present. Apparently, the electrostatic energy has surpassed the entropic scale, yet is still dominated by the ion-solvent binding energy.

In addition, it can be seen that the formation of the intermediate layer at potentials $\elpot>\SI{0.05}{\volt}$ in the solvation case shown in \cref{fig:profile_solv} increases the thickness of the EDL compared with the case without solvation. This property can be rationalised by the comparison of two electrolytes with the same ion size, once with solvation and once without: the solvation shell increases the effective size of the ion. As it is tightly bound to the central ion, the solvation shell can be considered part of a larger fundamental particle with an effective volume given by
\begin{equation}
    \label{eq:effective_volume}
    \pmv_\pm^{\ce{solv}}=\pmv_\pm+\cordnumb_\pm\pmv_{\solvent} = \pmvil(\pmvrel_\pm + \cordnumb_\pm\pmvrel_{\solventlabel}).
\end{equation}
Thus, if the solvent is not much smaller than the ions, the effective size of the solvated ion is significantly larger than the ``naked'' ion, $\pmv_\pm^{\ce{solv}}>\pmv_\pm$. The corresponding dilution parameter (bulk volume fraction of the solvated ions) is 
\begin{equation}
\label{eq:effective_dilution}
\dilution^{\ce{solv}}=\concbulk_{\ce{salt}}\left(\pmv_+^{\ce{solv}}|_{\ce{bulk}} + \pmv_-^{\ce{solv}}|_{\ce{bulk}} \right) = \dilution \left(1+\left[\cordnumb^{\ce{b}}_++\cordnumb^{\ce{b}}_-\right]\pmvrel_{\solventlabel}\right),
\end{equation}
where $\cordnumb_\pm^{\ce{b}}$ are the coordination numbers in the bulk. This indicates that, if we include solvation, the electrolyte behaves similar to a more concentrated electrolyte without solvation. Also, because the effective volume of the saturated ions increases, the saturation concentration of the solvated ions decreases, which in turn reduces the local charge density. Hence, before the electric field is strong enough to strip the ions of their solvation shells, the decreased charge density then requires a thicker EDL to screen the electrode. Consequently, the saturation of the stripped ions is also reached at a much higher potential (about \SI{0.6}{\volt} with solvation vs. \SI{0.1}{\volt} without). 

Altogether, we conclude that adding solvation interactions to our electrolyte model has two major effects. First, the solvent is pulled closer to the electrode, and, second, the EDL thickness increases.

\subsection{Influence of Dilution in non-interacting electrolytes}
\label{sec:analysis_nosolvation}

The goal of this section is to investigate the influence of dilution on the equilibrium structure of the EDL. In particular, we neglect solvation effects in this section by setting $\maxcoord_\pm$ to zero. This will allow the discrimination between effects due to electrolyte dilution from effects due to solvation, once we discuss the ``full'' model in \cref{sec:analysis_solvation}. 

\begin{figure}[tb]
\begin{center}
    \includegraphics[width=\columnwidth]{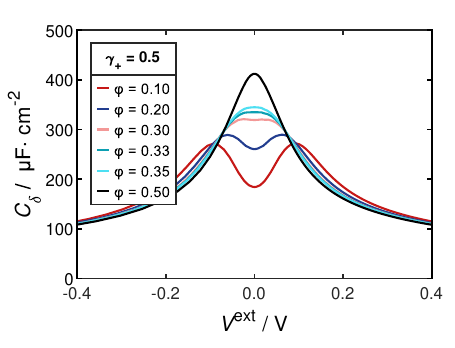}
    \caption{Differential capacitance $\diffcap$ versus applied potential $V^{\ce{ext}}$ of a symmetric electrolyte ($\pmvrel_\pm = 0.5 $) for different dilutions $\dilution$ without solvation.}
    \label{fig:no_solvation}
\end{center}
\end{figure}

First, we consider a fully ``symmetric'' electrolyte, where the ions and the solvent all have the same partial molar volume ($\pmvrel_\pm= \pmvrel_{\solvent} = 0.5$). \Cref{fig:no_solvation} shows the differential capacitance $\diffcap$ versus the external potential $V^{\ce{ext}}$ for this electrolyte at different values for the electrolyte dilution $\dilution$. For $\dilution=\num{0.1}$, i.e. at low salt concentration - or high dilution, the profile of $\diffcap$ exhibits two peaks (``camel'' shape), which are placed symmetrically around $V^{\ce{ext}}=\SI{0}{\volt}$. A slightly increased $\dilution=\num{0.2}$ raises the profile to slightly higher values, and dampens the central minimum . However, if the amount of salt in the electrolyte is increased above a critical value, which is roughly at $\dilution^{\ce{crit}} \approx 1/3$, the profile transitions towards having one peak (``bell'' shape)  at $V^{\ce{ext}}=0$. Further increasing the salt content $\dilution$ in the electrolyte, the profile remains bell-shaped, and the height of the central peak rises.

The transition of the DC profile from a ``camel'' shape (two peaks) to a ``bell'' shape (single peak) for symmetric electrolytes is well described in the literature, and has been studied using theoretical models based on mean-field theories,\cite{Kornyshev2007,Fedorov2014,Goodwin2020} and Gouy-Chapman-Stern models.\cite{Oldham2007} In particular, the critical value $\dilution^{\ce{crit}} = 1/3$ at which the transition occurs, was described analytically by Kornyshev.\cite{Kornyshev2007} Apparently, our model reproduces this finding for symmetric electrolytes. In the SI (\cref*{sec:KornyshevAnalytical}) we present an analytical discussion, where we show that our theory predicts the critical value $\dilution^{\ce{crit}}=1/3$.

\begin{figure}[tb]
    \begin{center}
    \includegraphics[width=\columnwidth]{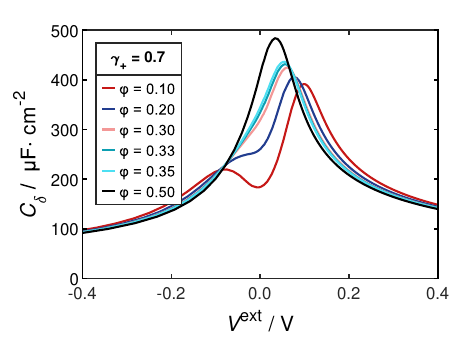}
    \caption{Differential capacitance $\diffcap$ versus applied potential $V^{\ce{ext}}$ of an asymmetric electrolyte with unequal ion sizes 
    for different dilutions $\dilution$ without solvation.}
    \label{fig:asymmetric_cap}
\end{center}
\end{figure}

Next, we discuss asymmetric electrolytes where the partial molar volumes of the cations and anions are unequal. \Cref{fig:asymmetric_cap} shows the DC profile as function of the applied potential $V^{\ce{ext}}$ for $\pmvrel_\cationlabel=\num{0.7}$ ($\pmvrel_\solventlabel=0.5$) and the same set of parameters $\dilution$ used in the symmetric case (see \cref{fig:no_solvation}). In the dilute regime ($\dilution=\num{0.1}$), the profile has a modified camel shape, where the two capacity peaks have different heights. In particular, the much taller ``right'' peak at positive potentials corresponds to the accumulation of anions, which are smaller than the cations. Furthermore, the central minimum is slightly shifted towards more positive potentials. When the dilution parameter is slightly increased to $\dilution=\num{0.2}$, we observe that the DC profile begins transitioning from camel to bell shape. For all higher values $\dilution$, the bell shape persists. Hence, the critical dilution is reduced for the asymmetric case to $\dilution^{\ce{crit}}\approx\num{0.2}$, as compared to $\dilution^{\ce{crit}}=1/3$ for symmetric electrolytes. Also, we observe that the single peak of the bell shaped profiles is shifted away from $V^{\ce{ext}} = \SI{0}{\volt}$ towards more positive potentials. This shift decreases with $\dilution$. 

Altogether, we conclude that the DC profile either has a camel shape or a bell shape, and that the transition between both depends on the specific sizes of the electrolyte species and on the dilution $\dilution$.  In particular, the result $\dilution^{\ce{crit}}=1/3$ is true only for symmetric electrolytes. As a general trend, we find that the larger the solvent, the higher the critical dilution is (see phase diagrams in \cref*{sec:phase_diag}). 

\subsection{Influence of Dilution when considering solvation}
\label{sec:analysis_solvation}

\begin{figure}[tb]
\begin{center}
    \includegraphics[width=\columnwidth]{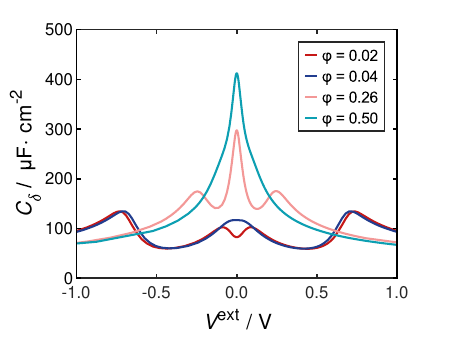}
    \caption{Differential capacitance $\diffcap$ versus applied potential $V^{\ce{ext}}$ of a symmetric electrolyte for different dilutions $\dilution$. }
    \label{fig:DC_with_solvation}
\end{center}    
\end{figure}

\begin{figure*}[tb]
\begin{center}
    \labelfig[width=\columnwidth]{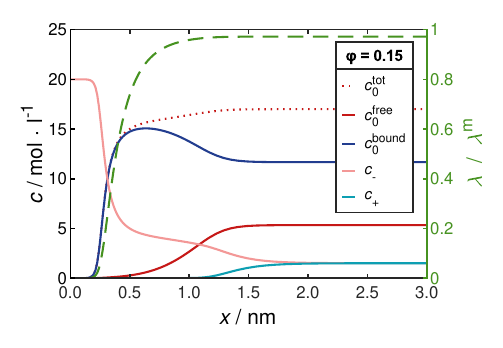}{fig:profile_dilute}
    \labelfig[width=\columnwidth]{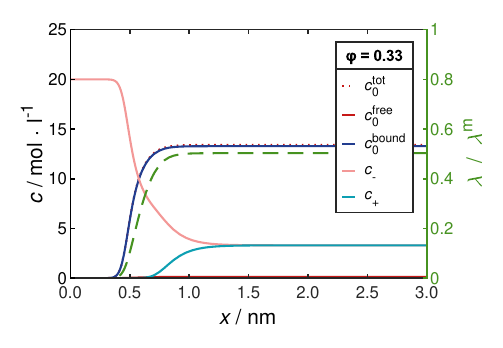}{fig:profile_conc}
    \caption{Concentration profiles of the free solvent molecules, the solvent molecules bound in a solvation shell and the ions at two different electrolyte dilutions. \Cref{fig:profile_dilute} shows results for a highly diluted electrolyte with $\dilution = 0.15$, and \cref{fig:profile_conc} shows results for $ \dilution= 0.33$.}
    \label{fig:diffcap_solvation_concs} 
    \end{center}
\end{figure*}

In this section, we include solvation effects into our numerical simulations of the EDL, and study its influence on the differential capacitance. For this purpose, we consider a symmetric electrolyte (equal size of the electrolyte species $\gammaup_\pm=\gammaup_{\solventlabel}=0.5$), and account for solvation effects by setting the maximal ion coordination numbers to $\maxcoord_\pm = 4$. Furthermore, we assume constant and equal binding energies between both ions and the solvent ($\tilde{E}_\pm = -4$).
In \cref{sec:discussion} we present a more detailed discussion of the complete parameter-landscape spanned by binding energy and size asymmetry. 

\Cref{fig:DC_with_solvation} shows numerical results for the differential capacitance $\diffcap$ versus applied potential for three different regimes of electrolyte dilution $\dilution$.  In the dilute case (here $\dilution = 0.02$), we observe four maxima, which are located symmetrically around the potential of zero charge  ($V^{\ce{pzc}}=\SI{0}{\volt}$). One pair of maxima occurs at lower potentials  (roughly at $\pm \SI{0.1}{\volt}$). These two peaks are reminiscent of the camel shaped profile of the case without dilution, see \cref{fig:no_solvation}. The two additional peaks at higher potentials (roughly at $\pm\SI{0.7}{\volt}$), however, are not present without solvation (see \cref*{fig:DC_nosolvation_wide}). This suggests, that the high-voltage peaks are a consequence of solvation. In a more concentrated electrolyte with $\dilution=\num{0.26}$, the DC profile exhibits three maxima. Apparently, the two inner peaks occurring when $\dilution=\num{0.02}$ are absorbed into a single central peak. This indicates a transition from camel shape to bell shape similarly observed  without solvation (see \cref{fig:no_solvation}). Simultaneously, the ``solvation peaks'' occurring at larger potentials are shifted towards the central peak with increasing $\dilution$. Finally, at $\dilution=\num{0.5}$, all peaks are absorbed into one central peak located at the pzc, where the height of the central peak increases with $\dilution$. 

The transition from four peaks (``camel''-shape) to three peaks (``bell''-shape) in the DC-plot occurs at about $\dilution=0.04$. This is significantly smaller than described in \cref{sec:analysis_nosolvation}, where no solvation was considered (there, $\dilution^{\ce{crit}}= 1/3$). This deviation can be explained by comparing the corresponding critical effective dilutions. The bulk coordination numbers can be assumed to be equal to $\maxcoord_\alpha$ in the dilute, strong solvation limit. We find that the simulation results for symmetric electrolytes shown in \cref{fig:no_solvation,fig:DC_with_solvation}, yield
\begin{equation}
    \label{eq:cameltobellsolvandnosolv}
    \dilution^{\ce{solv};\ce{crit}}\approx \dilution^{\ce{crit}}|_\textrm{no solvation}.
\end{equation}
\Cref{fig:DC_with_solvation} shows the critical dilution to be about \num{0.04}. Using \cref{eq:effective_dilution}, we can translate this to an effective value of $\dilution^{\ce {solv};\ce{crit}} \approx 0.2$. Accounting for the shift in the critical dilution from an increased effective size ratio of ions to solvent -- the increased effective salt volume (\cref{eq:effective_volume}) lowers $\pmvrel_\solventlabel$ to \num{0.1} -- we would also expect a critical dilution of 0.2 for a non-interacting electrolyte (see \cref*{fig:phase_diag_symmetric}).

Apparently, the effective dilution parameter $\dilution^{\ce{solv}}$ of the solvated ions, which accounts for the volume of the solvation shell, is the more proper descriptor for the DC profile than $\dilution$. 
This highlights that volumetric effects play a critical role in the formation of the EDL. 

Next, we illustrate how solvation affects the nano-structure of the EDL. \Cref{fig:diffcap_solvation_concs} shows the concentration profiles for two electrolyte configurations, which differ only by dilution. The cation (anion) species is shown in cyan (pink), while the solvent molecules are split into those bound to the ions in solvation shells (blue), and those which are not (red). Apparently, the positively charged electrode is screened by the accumulation of anions (pink curve) in both cases. \Cref{fig:profile_dilute} shows the numerical results for a diluted electrolyte ($\dilution \approx 0.15$), subject to an electrified interface with applied voltage $V^{\ce{ext}}=\SI{1}{\volt}$.  Here, the saturation layer described in \cref{sec:results_intro} spans a distance of  $L_{\ce{sat}}< \SI{1}{\nano\meter}$. The bulk region is characterised by the electroneutral configuration of the electrolyte where $\conc_+=\conc_-$, and starts after roughly \SI{2.5}{\nano\meter}. As can be seen, the unbound (red curve) and bound molecules (blue curve) of the neutral solvent species are pushed out of the EDL. The intermediate region extends roughly from \SI{0.8}{\nano\meter} up to \SI{2.5}{\nano\meter}. There, the concentration of the bound solvent molecules (blue curve) exhibits a maximum (roughly at \SI{0.6}{\nano\meter}), while the concentration of free solvent (red curve) is monotonically increasing moving away from the electrode. Beyond a distance of \SI{2}{\nano\meter} from the electrode, both reach their bulk values, where $\conc_{\solventlabel}^{\ce{free}}>\conc_{\solventlabel}^{\ce{bound}}$. The right y-axis shows the normalised coordination number $\tilde{\cordnumb}_-=\cordnumb_-/\maxcoord_-$ of the anions (dashed green curve). Apparently, the solvation shells are completely empty near the interface ($\lim_{x\to \num{0}}\tilde{\cordnumb}_-=0$). However, after roughly the width of the saturation layer $L_{\ce{sat}}$, the quantity $\tilde{\cordnumb}_-$ increases sharply with increasing distance from the electrode until it reaches it's bulk value at roughly \SI{1.0}{\nano\meter} (where the solvation shells of the ions are almost completely filled). 

\Cref{fig:profile_conc} shows numerical results for the species concentrations of a highly concentrated electrolyte, where $\dilution = 1/3$. Here, the screening of the positively charged interface by the anions ranges over roughly \SI{1}{\nano\meter}, after which the bulk phase of the electrolyte is reached. In contrast to the dilute case shown in \cref{fig:profile_dilute}), no intermediate region with a concentration maximum of the bound solvent molecules is visible. This is similar to the non-interacting case shown in \cref{sec:results_intro}, with the concentrations of the free and bound solvent molecules exhibiting more trivial profiles. In the bulk region of the electrolyte, we observe that $\conc_{\solventlabel}^{\ce{bound}}|_{\ce{bulk}}\gg\conc_{\solventlabel}^{\ce{free}}|_{\ce{bulk}}$. Hence, the solvent molecules in the bulk electrolyte are preferably bound in the solvation shell. This implies that in highly concentrated electrolytes most of the solvent molecules are coordinated with ions. Eventually, the amount of solvent becomes too low as to completely fill the solvation shells of the ions. This is evident by the bulk coordination number remaining below the theoretical limit $\cordnumb_\pm^{\ce{bulk}}<\maxcoord_\pm$. 

Next, we shall make use of the observations from above and discuss them from a more holistic perspective. Apparently, the DC profiles shown in \cref{fig:DC_with_solvation}, which involve the results of many individual EDL simulations (many different configurations of the electrolyte), draw a more cumulative picture as compared to the detailed results for the nanostructure of the EDL in one specific configuration shown in \cref{fig:diffcap_solvation_concs}. However, we can identify relations between the two sets of results and draw some conclusions for the behaviour of the system.

First, we conclude that there is a relation between the outer DC peaks in dilute electrolytes at larger potentials (see \cref{fig:DC_with_solvation}) and the sharp increase of the anion concentration towards their saturation threshold at a distance of $L_{\ce{sat}}\approx\SI{0.3}{\nano\meter}$ away from the interface in \cref{fig:profile_dilute}. This distance coincides with the complete stripping of the solvation shell. Thereby, the effective volume of the ions in the EDL is reduced to the volume of the ``naked'' ions, which allows for the ions to pack more densely in the EDL, and leads to a significant increase of the charge stored in the EDL. Because $C_\delta$ measures the rise of the EDL charge, this then results in the respective capacitance peaks at larger potentials. In the following, we refer to the potentials at which the outer DC peaks occur as the ``desolvation potential''. 

Furthermore, we conclude that the emergence of the intermediate region where the solvated ions saturate (see \cref{fig:diffcap_solvation_concs}), depends on the existence of an excess number of free (unbound) solvent molecules in the bulk of the electrolyte (see \cref*{sec:EDL_profile_study} for a more extensive study on EDL-profiles for different electrolytes). In turn, this implies that no desolvation peaks arise in the DC profile for higher values of $\dilution$, i.e. for more concentrated electrolytes (see \cref{fig:DC_with_solvation}). Accordingly, due to the removal of the intermediate layer, the interface is screened more efficiently for higher values of $\dilution$, and the width of the EDL decreases with increasing $\dilution$.

Finally, we show in \cref*{sec:EDL_profile_study} that the order in which the electrolyte species are pushed out of the EDL is independent of the electrolyte parameters (for a negative, i.e attractive, ion-solvent binding energy): first, the ions of the same charge as is applied to the interface are removed and second, at higher potentials, free solvent is displaced; third, the remaining solvated ions are stripped, and the bound solvents are pushed out of their solvation shells (desolvation of the counterions). However, if the applied voltage is not large enough to overcome the ion-solvent binding, there must not occur any desolvation at all.

\begin{figure}[tb]
    \begin{center}
    \includegraphics[width=\columnwidth]{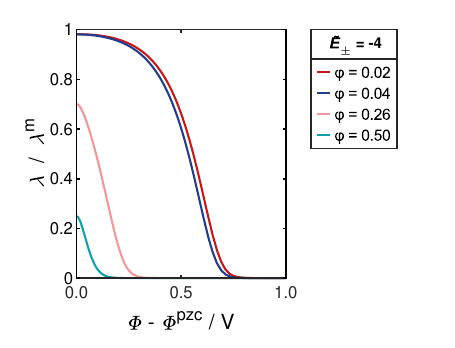}
    \caption{This plot shows the degree to which the solvation shells of the ions are filled at different points in the EDL and different bulk salt concentrations (indicated by $\dilution$). The x-axis represents the potential relative to the bulk potential (here $\elpot^{\ce pzc}$). At a potential of \SI{0}{\volt} the solvation shells of the bulk ions are displayed. The potential of \SI{1}{\volt} coincides with the electrolyte surface in this simulation, where a \SI{1}{\volt} potential difference between electrode and electrolyte is applied.}
    \label{fig:dilution}
    \end{center}
\end{figure}

Comparing with \cref{fig:DC_with_solvation}, we see that the outer peaks -- due to the desolvation of the ions -- are  increasingly shifted towards the origin for increasing $\dilution$. Eventually, at $\dilution = \num{0.5}$, the two outer peaks are absorbed into the central peak. This can be explained by the order in which the electrolyte species are displaced from the EDL. We found that the first species being pushed out of the EDL are ions of the same charge as the interface, followed by the free solvent species. Only once the desolvation potential is reached, is the bound solvent pushed out of the EDL. Thus, a reduction of the available free solvent (by increasing $\dilution$) decreases the number of counter-ions that can be exchanged for the solvent molecules. This inferior screening of the electrode leads to a greater driving force removing the solvation shell from the ions. Hence, the desolvation potential decreases for highly concentrated electrolytes, even when the binding energy remains the same. Additionally, if the solvation shells are only partially filled even in the bulk, the size reduction from stripping the bound solvent is not as significant. Because the increase in charge density at the desolvation potential is less significant, the desolvation peaks become less pronounced for higher $\dilution$. Both of these factors, the location of the peak closer to the large central peak and reduced charge gain, make the secondary peak disappear in highly concentrated electrolytes, indicating an effectively weaker ion-solvent bond. 

\Cref{fig:dilution} shows the effect of different dilutions on the solvation shell occupancy. We again consider a symmetric electrolyte (where $\pmvrel_\pm=\pmvrel_{\solvent}=\num{0.5}$), and assume that $\ndbinding=-\num{4}$ and $\maxcoord_\pm=\num{4}$. 
The plot is shown against the electric potential in the electrolyte $\elpot-V^{\ce{pzc}}$, as  there is a direct relationship between the space coordinate and the local potential (discussed in \cref{sec:comp}). This was chosen to better highlight the shift of the desolvation potential with $\dilution$. $\elpot-V^{\ce{pzc}}=\SI{0}{\volt}$ corresponds to the bulk electrolyte region. For small dilutions, up to $\dilution=\num{.15}$, the solvation shells are completely filled in the bulk. They also remain fully filled in the EDL up to roughly $\elpot-V^{\ce{pzc}}=\SI{0.5}{\volt}$. Beyond this desolvation potential, however, the solvation shells are quickly stripped and the normalised coordination number $\tilde{\cordnumb}_\alpha$ decreases exponentially to zero. Beginning with $\dilution=\num{0.15}$, the filling of the solvation shells in the bulk decreases with increasing values $\dilution$. In the EDL, the coordination numbers decay exponentially towards zero, i.e. empty solvation shells. Hence, for higher amounts of salt in the electrolyte, the solvation shells in the bulk are only partially filled. Apparently, if we increase $\dilution$ above roughly 0.2, there aren't enough solvent molecules to fill the solvation shells, so there have to remain empty sites. Notably, at high salt concentrations barely any free solvent remains (see also \cref{fig:diffcap_solvation_concs}).

Summarising, for highly concentrated electrolytes where $\dilution\to\num{1}$, we find that $\xparam\to\num{0}$. \Cref{eq:extremal_condition_coordnumb} then implies that the solvent coordination number converges to zero, i.e. empty solvation shells, regardless of the binding energy $\ndbinding$. Altogether, an increased dilution (reduced $\dilution$) of the electrolyte corresponds to an entropic stabilization of the solvation shell, (similar to $\ndbinding$, which corresponds to an energetic stabilization of the solvation shell).

\begin{figure*}[ht]
    \begin{center}
    \labelfig[width=8.5cm]{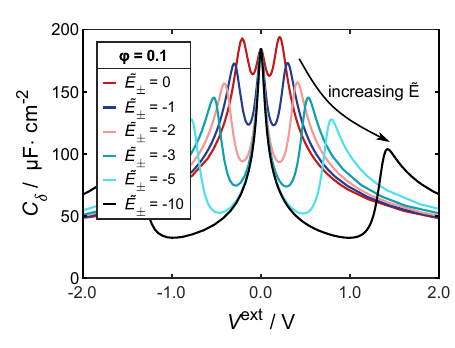}{fig:binding_energy_DC}
    \labelfig[width=4.25cm]{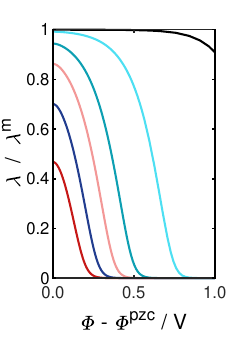}{fig:binding_energy_lambda}
    \labelfig[width=4.25cm]{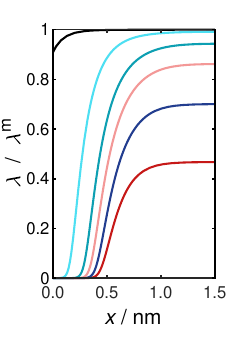}{fig:binding_energy_lambda_x}
    \caption{Influence of binding energy $\ndbinding_\pm$ on the solvation of a symmetric, diluted ($\dilution=\num{0.1}$) electrolyte. \Cref{fig:binding_energy_DC} shows the differential capacitance $\diffcap$ versus applied potential $V^{\ce{ext}}$. \Cref{fig:binding_energy_lambda} shows the filling degree $\tilde{\cordnumb}_\pm$ of the solvation shells versus the local electric potential $\elpot-\elpot^{\ce{pzc}}$ in the electrolyte. For reference, in \cref{fig:binding_energy_lambda_x} the same data as in \cref{fig:binding_energy_lambda} is plotted against the space variable. As $x$ measures the distance to the electrode, the point at $x=0$ corresponds with $\elpot-\elpot^{\ce{pzc}}= \SI{1}{\volt}$, the potential difference applied between the bulk electrolyte and the electrode.}
    \label{fig:binding_energy}
    \end{center}
\end{figure*}

\section{Discussion}
\label{sec:discussion}

In this section, we discuss the influence of the electrolyte parameters on the solvation of the ions in the EDL.  
First, in \cref{sec:discussion_bindingE}, we focus on the binding energy. Second, in \cref{sec:discussion_sizeasymmetry}, we examine the influence of ion-size asymmetry. Next, in \cref{sec:validation} we qualitatively reproduce experimental results for a real electrolyte. Finally, in \cref{sec:pzc_discussion}, we touch on our choice of normalisation for the potential, the potential of zero charge, and discuss its relationship with the capacitance minimum.

\subsection{Influence of Binding Energy}
\label{sec:discussion_bindingE}

As discussed in \cref{sec:model}, our continuum model introduces two parameters ($\ndbinding_\alpha$ and $\cordnumb_\alpha$) to account for solvation. In this section, we focus on the normalised ion-solvent binding energy $\ndbinding_\alpha  = E_\alpha/k_{\ce{B}}T$ (at room temperature $T=\SI{298}{\kelvin}$) and study its influence on the EDL solvation.

\Cref{fig:binding_energy_DC} illustrates the differential capacitance of a symmetric ($\pmvrel_\alpha=\num{0.5}$), dilute ($\dilution=\num{0.1}$) electrolyte as function of the external potential, for six different binding energies. Here, we set $\maxcoord_\alpha=\num{4}$. We emphasise that solvation occurs, indicated by non-vanishing coordination numbers ($\cordnumb_\alpha\neq0$), even if the binding energy is set to zero ($\ndbinding_\alpha=\num{0}$). For all binding energies, the profiles exhibit three peaks, with one peak located at $V^{\ce{ext}}=\SI{0}{\volt}$ and two secondary peaks located at higher (negative and positive) potentials. The choice of $\dilution=\num{0.1}$ leads to a ``bell''-shape (i.e. single peak) in the DC profile with respect to the ``inner'' peaks at lower potentials (see \cref{fig:no_solvation}). This central peak is unaffected by changes in the binding energy, which is in accordance with our interpretation from above, where we concluded that ``inner'' peaks are only due to effects of ion size and dilution (see description in \cref{sec:analysis_nosolvation}). The outer peaks, however, are strictly dependent on the binding energy and are shifted towards higher (negative and positive) potentials with increasing binding energy. 

\Cref{fig:binding_energy_lambda} illustrates the normalised coordination number $\tilde{\cordnumb}_\alpha$ as function of the electric potential $\elpot$. We normalise the potential with respect to the potential of zero charge (pzc) $\elpot^{\ce{pzc}}$, which in our simulation is set to be the bulk potential. This is a consequence of choosing an inert electrode -- if we hold the electrode at the bulk potential, no charge accumulates in the EDL. The value of $\tilde{\cordnumb}_\alpha$ at $\elpot=\elpot^{\ce{pzc}}$ correspond to the coordination numbers in the bulk, whereas the results for $\tilde{\cordnumb}_\alpha$ at increased potentials indicate the filling of the solvation shells in different regions of the EDL. For vanishing binding energy, the solvation shells in the bulk are filled to only 50\%. The degree of filling depends on the binding energy, where higher binding energy increases the solvent coordination in the bulk. The bulk solvation shells are nearly completely filled from $\ndbinding = -5$, while at \SI{1}{\volt} the binding energy required to fill the shells is larger than $\ndbinding = -10$. At higher potentials, the normalised solvent coordination number $\tilde{\cordnumb}_\alpha$ decays exponentially towards complete desolvation of the ions ($\tilde{\cordnumb}_\alpha=\num{0}$). 

Apparently, the desolvation potential (the potential where $\tilde{\cordnumb}_\alpha\approx\num{0}$) increases with increasing binding energy. A quantitative analysis of \cref{fig:binding_energy_DC,fig:binding_energy_lambda} reveals that the voltage $V^{\ce{ext}}$, at which the outer capacitance peaks are located, always correlate to a filling of the solvation shells of around 3\%, i.e. almost empty shells. This again confirms our conclusion that the outer peaks arise due to the stripping of the solvation shell and are located exactly at the desolvation potential ($\tilde{\cordnumb}_\alpha\approx\num{0}$).  The increase in desolvation potential with increasing binding energy shows that higher binding energies lead to more stable solvation shells, which are more resistant to being stripped. Thus, the local potential must be high enough as to overcompensate the strong ion binding. 

Finally, we briefly discuss our observation from \cref{fig:binding_energy_lambda}, that $\ndbinding_\pm$ also affects the solvation shells in the bulk (i.e. at $\elpot-\elpot^{\ce{pzc}}=\SI{0}{\volt}$). Our statistical theory presented in \cref{sec:model} yields an implicit relation between the filling of the solvation shells and the binding energy via $\tilde{\cordnumb}_\alpha = \cordnumb_\alpha/\maxcoord_\alpha = \xparam /(\xparam + \exp(-|\ndbinding_\alpha|))$, 
where $x(\tilde{\cordnumb}_\beta)=\conc_{\solventlabel}^{\ce{free}}/\conc^{\ce{free}}$ is always smaller than unity ($\xparam\leq 1$). Thus for large (negative) binding energies $|\ndbinding|\gg\num{1}$, the coordination number converges to unity, $\tilde{\cordnumb}_\alpha\to\num{1}$. In this limit, the solvation shells become very stable and can hardly be stripped in the EDL. In contrast, for $\ndbinding\to \num{0}$, the coordination number can be approximated by $\tilde{\cordnumb}_\alpha = \xparam/(1+\xparam)$. Thus, for highly diluted electrolytes and $\ndbinding=\num{0}$, we find that $\xparam\to \num{1}$. \Cref{eq:extremal_condition_coordnumb} then implies $\tilde{\cordnumb}_\alpha\to 1/2$, which matches exactly the numerical results shown in \cref{fig:binding_energy_lambda}.

\begin{figure*}[!htb]
    \begin{center}
    \labelfig[width=\columnwidth]{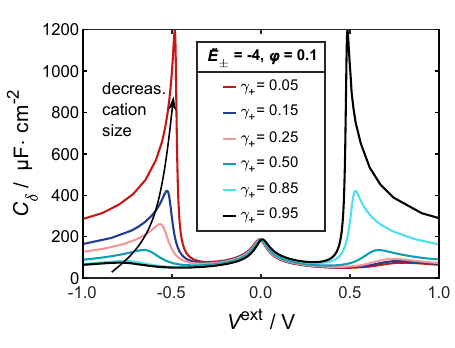}{fig:ion_asymmetry_DC}
    \labelfig[width=\columnwidth]{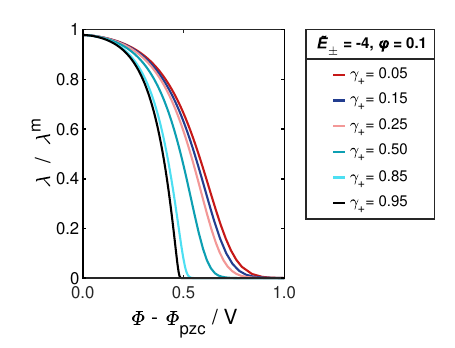}{fig:ion_asymmetry_lambda}
    \caption{Influence of ion-size asymmetry $\pmvrel_\pm$ on the solvation of a diluted ($\dilution=\num{0.1}$) electrolyte. \Cref{fig:ion_asymmetry_DC} shows the differential capacitance $\diffcap$ versus applied potential $V^{\ce{ext}}$. \Cref{fig:ion_asymmetry_lambda} shows the filling degree $\tilde{\cordnumb}_\pm$ of the solvation shells versus the local electric potential $\elpot-\elpot^{\ce{pzc}}$ in the electrolyte.}
    \label{fig:ion_asymmetry}
    \end{center}
\end{figure*}

\subsection{Influence of the size asymmetry on the differential capacitance}
\label{sec:discussion_sizeasymmetry}

In this section, we study the influence of the size asymmetry $\pmvrel_+$ between the cations and anions on the solvation in the EDL (recall that $\pmvrel_\alpha=\pmv_\alpha/\pmvil$ and $1 = \pmvrel_+ + \pmvrel_-$). When varying $\pmvrel_+$, we keep $\pmvil$ and the solvent size constant, such that $\dilution$ remains unaffected from the variation of $\pmvrel_\pm$. Note that $\pmvrel_+>\num{0.5}$ implies that the cations are larger than the anions.

\Cref{fig:ion_asymmetry_DC} shows the differential capacitance as function of the external potential for six different size asymmetries $\pmvrel_+$, where $\dilution=\num{0.1}$, $\maxcoord_\alpha=\num{4}$, and $\ndbinding=-\num{4}$. For all size asymmetries, the DC profiles exhibit three different peaks (this is in accordance with results from above for $\ndbinding=-\num{4}$ and $\dilution = \num{0.2}$ and thus corresponds to a bell shape). The main peak at $V^{\ce{ext}} = \SI{0}{\volt}$ is minimally shifted to the right with increasing $\pmvrel_+$. This can be attributed to the fact that $\dilution$ remains constant. In contrast, the amplitudes of the secondary peaks (indicating the desolvation potential) increase with decreasing size of the screening species (cations for negative potentials $V^{\ce{ext}}$, and anions for positive potentials $V^{\ce{ext}}$). The smallest cation shown here, at $\pmvrel_{\cationlabel}=0.05 $, gives a very high desolvation peak at around $V^\textrm{ext}=\SI{-0.5}{\volt}$. We explain this behaviour as follows. For very small ions ($\pmvrel_\pm \ll \pmvrel_{\solvent}$),  there exists a significant size difference between a desolvated ion and a solvated ion (see also the discussion on effective size in \cref{sec:analysis_solvation}). Thus, when the solvation shell of a small ion gets stripped and the newly formed free solvent molecules are quickly pushed out of the EDL, the ion concentration can rise sharply by displacing the solvent. Thus, the EDL charge is also abruptly increased. In effect, this results in the prominent DC peak.  

\Cref{fig:ion_asymmetry_lambda} shows that the desolvation potential decreases with decreasing size of the screening ion. Therefore, the solvation shells of smaller ions are, effectively, less stable than that of larger ions (with equal binding energy). In contrast, the bulk solvation is unaffected by size asymmetry of the salt.

\begin{figure*}
    \begin{center}
    \labelfig[width=\columnwidth]{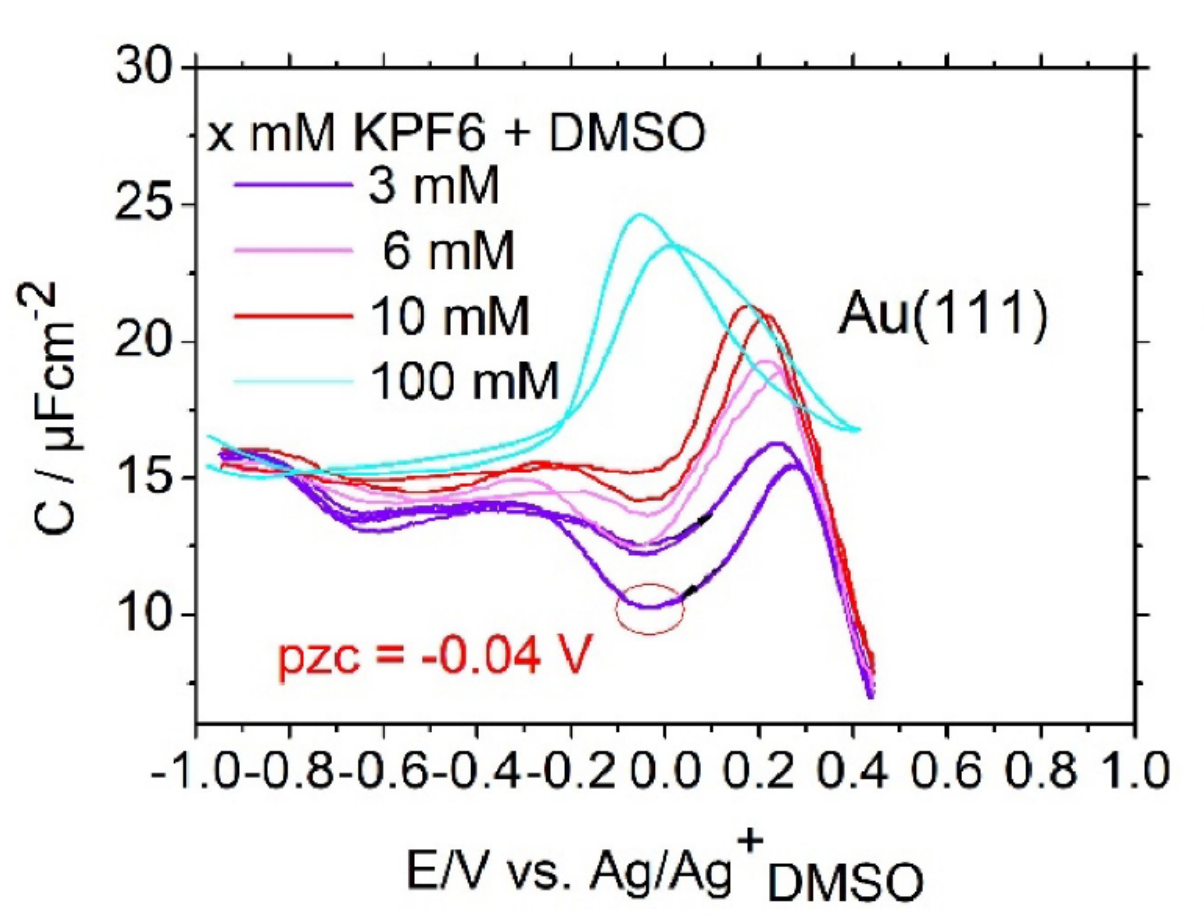}{fig:Shatla_exp}
    \labelfig[width=\columnwidth]{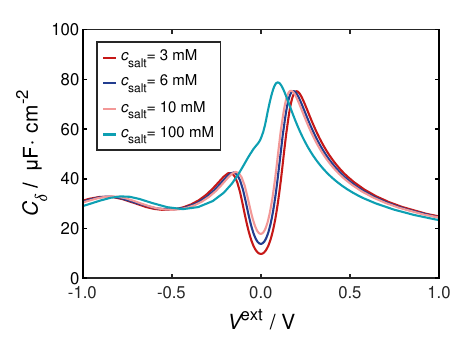}{fig:Shatla_simulation}
    \caption{Differential capacitance $\diffcap$ 
    of a \ce{KPF_6} salt in DMSO electrolyte with different mixing ratios. \Cref{fig:Shatla_exp} shows experimental results as presented in Ref.~\citenum{Shatla2021}. \Cref{fig:Shatla_simulation} shows numerical results as obtained from computer simulations.}
    \label{fig:validation}
    \end{center}
\end{figure*}

\subsection{Validation: Comparison With Experiment}
\label{sec:validation}

In this section, we validate our simulation results in comparison with experimental data from the literature. For this purpose, we focus on results presented by Shatla et al. for their measurement of the differential capacitance, see Ref.~\citenum{Shatla2021}. In this publication, the authors describe the electrolyte mixture composed of a \ce{KPF_6} salt in a DMSO solvent on a Gold electrode (Au(111)), and obtain the differential capacitance from impedance measurements using a three electrode setup. In contrast to many publications in the literature, the authors perform measurements over a wide potential window and  present results which extend beyond the narrow potential window around the potential of zero charge (pzc). Such a wide potential window is crucial for validating our model predictions, in particular with respect to the emergence of secondary peaks in the differential capacitance at larger potentials (desolvation potentials due to the stripping of solvation shells). 

\Cref{fig:Shatla_exp} illustrates experimental results for the differential capacitance for different salt / solvent ratios. The pzc at roughly $V^{\ce{pzc}}=-\SI{40}{\milli\volt}$ is indicated by a red circle. For electrolyte compositions with small salt concentrations, the DC profiles exhibit three peaks. The peaks are not arranged symmetrically, but rather exhibit different magnitudes and relative potentials. With increasing salt amount, all peaks are shifted toward the pzc. Also, the magnitudes of the peaks increase. Eventually, when the salt amount gets large enough (here for $\conc^{\ce{bulk}} \geq \SI{100}{\milli\molar}$), the central peaks are absorbed into one peak.  

We use these results for the validation of our theoretical approach. For this purpose, we parameterised an electrolyte in accordance with Ref.~\citenum{Shatla2021} (see \cref*{sec:param_validation} for details of the parameterisation) and performed numerical simulations. In particular, we neglect solvation of anions via neutral solvent molecules and set $\ndbinding_-=\num{0}$. This assumption can be motivated by the observation that in common battery electrolytes, the counter ions \ce{PF_6^-} and \ce{TFSI^-} are both only weakly coordinating in carbonate solvents. \cite{Chen2019,vonWald2015}
\Cref{fig:Shatla_simulation} shows our numerical results for this system at the same salt amounts as in the  the experimental results. As opposed to the experimental data, in our simulation the pzc is set at $V^{\ce{pzc}}=\SI{0}{\volt}$. For salt concentrations up to $\conc^{\ce{bulk}}_{\ce{salt}}=\SI{10}{\milli\molar}$, the DC profiles show three different peaks, where two peaks lie in the ``negative'' branch ($V^{\ce{ext}}-V^{\ce{pzc}}<0$), and only one peak (with large magnitude) lies in the ``positive'' branch ($V^{\ce{ext}}-V^{\ce{pzc}}>0$). We emphasise that this is a consequence of our assumption that the anions do not to participate in solvation effects. The inner peaks of the negative branch (occurring at smaller negative potentials) can be attributed to the effect of dilution, whereas the outer peaks at larger potentials arise when the solvation shell is stripped at high enough potentials (see \cref{sec:analysis_nosolvation,sec:analysis_solvation}). The transition from bell to camel shape at higher salt concentrations can still be observed. 

Overall, the simulation results are qualitatively in good agreement with the experimental results. The occurrence of three peaks in the experimental results, as well as the location of these capacitance peaks is well reproduced by the numerical data. However, the height of the DC peaks differ by a factor of about 4 between the experimental results ($C_\delta^{\ce{peak}}\leq\SI{20}{\micro\farad\per\centi\meter\squared}$) and the numerical results ($C_\delta^{\ce{peak}}\leq\SI{80}{\micro\farad\per\centi\meter\squared}$).

These discrepancies are influenced by the strong dependence of the differential capacitance on the size of the ions. For example, if the simulation would be done with larger ion sizes, this would reduce the overall capacitance values. Here, we parameterised our electrolytes based on literature data (see \cref*{sec:param_validation}). In particular, the values for the partial molar volumes of the electrolyte species $\pmv_\alpha$ were obtained by a rough approximation from the crystalline ionic radii of the ions. However, to accurately model the size of the ions in solution, results obtained from molecular dynamics or DFT considering solvation shells would be more suitable. Unfortunately, these are difficult to obtain. 

Furthermore, we neglect non-local correlations between the ions and the solvent. Recently, we have shown that such interactions can have a significant influence on the structure of the EDL, and may lead to long-ranged charge oscillations.\cite{Hoffmann2018,schammer2021role} By extending the free energy density shown in \cref{eq:bulk_freeenergy_all}, non-local species interactions can easily be incorporated into our theory of solvation. We expect that this modified theory would yield a more heterogeneous structure of the EDL, and broaden the width of the EDL to several ion diameters. Arguably, this would reduce the magnitude of the differential capacitance, and improve the quantitative comparison with experimental data. Furthermore, we also assume a constant dielectric parameter in this publication. This description could be refined by including a dependence of the dielectric parameter on the species concentration. In particular, the dielectric parameter should decrease with decreasing solvent concentration. A reduced dielectric constant inside the highly concentrated EDL, in turn, also lowers the magnitudes of the differential capacitance. However, including such non-local interactions into our description goes beyond the scope of the present paper.

\subsection{Potential of zero charge}
\label{sec:pzc_discussion}

Finally, we briefly comment on the relation of our description for the differential capacitance and the potential of zero charge (pzc). There exist different definitions for the pzc in the literature. Here, we follow the convention used by Shatla and Landstorfer for highly diluted electrolytes,\cite{Shatla2021} where the pzc is characterised by the voltage where the DC profile has its central capacitance minimum. The pzc of a given system depends both on the properties of the electrolyte, as well as on the properties of the electrode.\cite{Huang2021} Our model focuses on the electrolyte and does not yet include properties of the electrode (which is considered inert). In our simulations, for dilute solutions, the central minimum occurs always at $V^{\ce{ext}}-V^{\ce{pzc}}=\SI{0}{\volt}$ (see \cref{sec:results_difcap}). Hence, the minimum of the camel-shaped DC profile for highly diluted electrolytes always coincides with the pzc, and, thus, our model reproduces the definition for the pzc. However, if we stray away from infinite dilution by increasing $\dilution$, we observe that the ion asymmetry and solvent size has an effect on the location of the minimum. In general, if the ions are of equal size, the minimum always coincides with the pzc, also for higher values of $\dilution$ (see \cref{fig:no_solvation,fig:DC_with_solvation,fig:dilution}). In contrast, for asymmetric ions, the minimum of the DC shifts to higher (positive) potentials for a larger anion, and in the negative direction for a larger cation (see \cref{fig:asymmetric_cap}). As we show in the SI (see \cref*{sec:asymmetry_pzc}), this effect gets more pronounced with increasing asymmetry between the ion species, with decreasing solvent size and with increasing salt concentration (increasing $\dilution$). Thus, the capacitance minimum can deviate from the pzc depending on electrolyte properties like dilution or ion-size-asymmetry. 

\section{Conclusion}
\label{sec:conclusion}

In this work, we present and discuss a continuum model for liquid electrolytes which incorporates solvation effects. This model is based on our framework for highly concentrated liquid electrolytes.\cite{schammer2020theory} The solvation of ions by solvent molecules is captured via including two additional terms to the free energy of the system, which describe entropy reduction due to solvent immobilisation and the binding energy between the ions and the solvent. In particular, our model for solvation needs only two additional parameters. The resulting description applies to a wide variety of electrolyte mixtures, ranging from dilute electrolytes to pure salts, and allows to study their dynamical behaviour. Here, we apply our general theory to electrolytes composed of a salt mixed with a neutral solvent, and focus on the equilibrium structure of the EDL. This is conveniently analysed using stationary solutions of our transport model. Balancing the electrochemical forces in the EDL makes it possible to directly obtain the local structure from the bulk electrolyte properties.

We discuss the influence of solvation and dilution on the differential capacitance, and validate our model predictions in comparison with experimental results for \ce{KPF_6} in \ce{DMSO} electrolytes. Our model reproduces well-known dilution effects discussed in the literature, e.g. the transition from a camel-shaped profile of the differential capacitance (DC) to a bell-shaped profile with increasing salt concentration. However, our model goes beyond current theories from the literature and draws a richer picture, as it predicts the emergence of additional peaks in the DC profile occurring at higher potentials. These peaks are due to solvation effects, and can be attributed to the complete stripping of the solvent molecules from the solvation shell of the ions in the EDL. Thus, our description rationalises the DC profile, and separates effects of dilution from effects of solvation. 

Our model also enables examination of a charged ``solvent''-species. For example, novel electrolyte mixtures composed of two salts with a common anion have recently been discussed in the literature as promising candidates for next generation batteries.\cite{doi:10.1021/acs.jpcb.1c05546}

In addition, we present a detailed discussion how the parameters influence the EDL solvation. Our analysis reveals that the EDL structure of the electrolyte depends on the size-asymmetry of the ions and the solvent molecules, the dilution of the electrolyte, and on the binding energy between the ions and the solvent molecules. 

Our continuum model is highly applicable to different electrolyte systems and and thus constitutes a promising starting point for future research. Because it is based on modelling the free energy, it can easily be supplemented by additional effects and interactions.

Our approach describes the electrolyte evolution via time-dependent transport equations, so dynamic transport can also be accurately described within the framework described in this manuscript. To the best knowledge of the authors, our approach therefore constitutes the first non-equilibrium solvation theory on the continuum scale.
This allows for studying the influence of currents on the
evolution of non-equilibrium EDL structures including solvation
effects. However, this goes beyond the scope of this manuscript.

%% file: tikzfileNumericalScheme.tex
\fontfamily{lmss}

\begin{tikzpicture}[scale=1.5]
    \draw (-4,-2) rectangle (4,2);
    \fill [left color = black!10!white, right color= black!40!white](4,-2) rectangle (4.5,2);
    \fill [black!40!white,overlay] (-5,-2) rectangle (-4,2);
    
    \node at (-4,-2) [below] {0};
    \node at (4,-2) [below] {$x^\textsf{b}$};
    \node[font = \large] at (0,-2.2) {1D simulation, $x = $ distance from electrode / nm};
    \node at (-4,0) [above, rotate=90] {\large\bfseries electrode};
    \node at (4,0) [below, rotate=90] {\large\bfseries bulk};
    
    \draw[<->,purple!70!blue,line width=1pt] (-4.4,-1) to (-3.6,-1);
    \node at (-4.15,-1.1) [below,purple!70!blue,font=\large] {$\boldsymbol J, \boldsymbol N_\alpha = 0$};
    \node at (3.8,1.4) [purple!70!blue,font=\large,align=left] {bulk \\charge $ = 0$\\density };
    \node at (-0.5,1.5) [purple!70!blue,font=\large,align=center] {\bfseries stationary state \\ $ \boldsymbol\nabla\cdot\boldsymbol J = \boldsymbol\nabla\cdot\boldsymbol N_\alpha = 0$};
    
    \node at (-4,-1.7) [right,blue!60!white, font=\large] {electrode potential $\Phi^\textsf{el}$};
    \node at (4,-1.7) [left,blue!60!white, font=\large] {bulk potential $\Phi^\textsf{bulk} = 0$};
    \draw[<-,blue!60!white] (-1.2,-1.7) to (1.1,-1.7) node [above, font=\large] at (0,-1.7) {applied voltage $V^\textsf{ext}$};
    
    \begin{scope}[shift={(1.75,0.8)}]
        \filldraw[blue!50!black, draw=black] (0,0) circle (0.3);
        \fill[blue!30!white, opacity=0.6] (0,0) circle (0.2);
        \node at (0,0) {\large\textcolor{white}{$\boldsymbol -$}};
        
        \node at (-0.4,-0.4) {\setchemfig{atom style={scale=0.6}, atom sep=0.35, bond offset = 0.35}\chemfig{[:155]*5(--O-(=O)-O-)}};
        \node at (0.4,-0.4) {\setchemfig{atom style={scale=0.6}, atom sep=0.35, bond offset = 0.35}\chemfig{[:-115]*5(--O-(=O)-O-)}};
        \node at (-0.4,0.4) {\setchemfig{atom style={scale=0.6}, atom sep=0.35, bond offset = 0.35}\chemfig{[:65]*5(--O-(=O)-O-)}};
        \node at (0.4,0.4) {\setchemfig{atom style={scale=0.6}, atom sep=0.35, bond offset = 0.35}\chemfig{[:-25]*5(--O-(=O)-O-)}};
        
        \draw[blue, dashed] (-0.4,-0.4) circle (0.25);
        \draw[blue, dashed] (0.4,-0.4) circle (0.25);
        \draw[blue, dashed] (-0.4,0.4) circle (0.25);
        \draw[blue, dashed] (0.4,0.4) circle (0.25);
        
        \draw[blue, dashed] (0,0) circle (0.816);
        
        \node at (-1.05,-0.15) {\setchemfig{atom style={scale=0.6}, atom sep=0.35, bond offset = 0.35}\chemfig{[:10]*5(--O-(=O)-O-)}};
    \end{scope}
    
    \begin{scope}[shift={(3.1,-0.4)}]
        \filldraw[red!70!black, draw=black] (0,0) circle (0.3);
        \fill[red!50!white, opacity=0.6] (0,0) circle (0.2);
        \node at (0,0) {\large\textcolor{white}{$\boldsymbol +$}};
        
        \node at (-0.4,-0.4) {\setchemfig{atom style={scale=0.6}, atom sep=0.35, bond offset = 0.35}\chemfig{[:-25]*5(--O-(=O)-O-)}};
        \node at (0.4,-0.4) {\setchemfig{atom style={scale=0.6}, atom sep=0.35, bond offset = 0.35}\chemfig{[:65]*5(--O-(=O)-O-)}};
        \node at (-0.4,0.4) {\setchemfig{atom style={scale=0.6}, atom sep=0.35, bond offset = 0.35}\chemfig{[:-115]*5(--O-(=O)-O-)}};
        \node at (0.4,0.4) {\setchemfig{atom style={scale=0.6}, atom sep=0.35, bond offset = 0.35}\chemfig{[:155]*5(--O-(=O)-O-)}};
        
        \draw[blue, dashed] (-0.4,-0.4) circle (0.25);
        \draw[blue, dashed] (0.4,-0.4) circle (0.25);
        \draw[blue, dashed] (-0.4,0.4) circle (0.25);
        \draw[blue, dashed] (0.4,0.4) circle (0.25);
        
        \draw[blue, dashed] (0,0) circle (0.816);
        
        \node at (0,1.2) {\setchemfig{atom style={scale=0.6}, atom sep=0.35, bond offset = 0.35}\chemfig{[:-70]*5(--O-(=O)-O-)}};
        \node at (-1.4,-0.3) {\setchemfig{atom style={scale=0.6}, atom sep=0.35, bond offset = 0.35}\chemfig{[:140]*5(--O-(=O)-O-)}};
        
    \end{scope}

    \begin{scope}[shift={(-1.3,0)}]
        \filldraw[red!70!black, draw=black] (0,0) circle (0.3);
        \fill[red!50!white, opacity=0.6] (0,0) circle (0.2);
        \node at (0,0) {\large\textcolor{white}{$\boldsymbol +$}};
        
        \node at (-0.4,-0.4) {\setchemfig{atom style={scale=0.6}, atom sep=0.35, bond offset = 0.35}\chemfig{[:-25]*5(--O-(=O)-O-)}};
        \node at (0.4,-0.4) {\setchemfig{atom style={scale=0.6}, atom sep=0.35, bond offset = 0.35}\chemfig{[:65]*5(--O-(=O)-O-)}};
        \node at (-0.4,0.4) {\setchemfig{atom style={scale=0.6}, atom sep=0.35, bond offset = 0.35}\chemfig{[:-115]*5(--O-(=O)-O-)}};
        \node at (0.4,0.4) {\setchemfig{atom style={scale=0.6}, atom sep=0.35, bond offset = 0.35}\chemfig{[:155]*5(--O-(=O)-O-)}};
        
        \draw[blue, dashed] (-0.4,-0.4) circle (0.25);
        \draw[blue, dashed] (0.4,-0.4) circle (0.25);
        \draw[blue, dashed] (-0.4,0.4) circle (0.25);
        \draw[blue, dashed] (0.4,0.4) circle (0.25);
        
        \draw[blue, dashed] (0,0) circle (0.816);
    \end{scope}
    
    \begin{scope}[shift={(0.3,-0.4)}]
        \filldraw[red!70!black, draw=black] (0,0) circle (0.3);
        \fill[red!50!white, opacity=0.6] (0,0) circle (0.2);
        \node at (0,0) {\large\textcolor{white}{$\boldsymbol +$}};
        
        \node at (-0.4,-0.4) {\setchemfig{atom style={scale=0.6}, atom sep=0.35, bond offset = 0.35}\chemfig{[:-25]*5(--O-(=O)-O-)}};
        \node at (0.4,-0.4) {\setchemfig{atom style={scale=0.6}, atom sep=0.35, bond offset = 0.35}\chemfig{[:65]*5(--O-(=O)-O-)}};
        \node at (-0.4,0.4) {\setchemfig{atom style={scale=0.6}, atom sep=0.35, bond offset = 0.35}\chemfig{[:-115]*5(--O-(=O)-O-)}};
        \node at (0.4,0.4) {\setchemfig{atom style={scale=0.6}, atom sep=0.35, bond offset = 0.35}\chemfig{[:155]*5(--O-(=O)-O-)}};
        
        \draw[blue, dashed] (-0.4,-0.4) circle (0.25);
        \draw[blue, dashed] (0.4,-0.4) circle (0.25);
        \draw[blue, dashed] (-0.4,0.4) circle (0.25);
        \draw[blue, dashed] (0.4,0.4) circle (0.25);
        
        \draw[blue, dashed] (0,0) circle (0.816);
    \end{scope}
    
    
    \begin{scope}[shift={(-2.75,0.75)}]
        \filldraw[red!70!black, draw=black] (0,0) circle (0.3);
        \fill[red!50!white, opacity=0.6] (0,0) circle (0.2);
        \node at (0,0) {\large\textcolor{white}{$\boldsymbol +$}};
        
        \node at (0.4,-0.4) {\setchemfig{atom style={scale=0.6}, atom sep=0.35, bond offset = 0.35}\chemfig{[:65]*5(--O-(=O)-O-)}};
        \node at (0.4,0.4) {\setchemfig{atom style={scale=0.6}, atom sep=0.35, bond offset = 0.35}\chemfig{[:155]*5(--O-(=O)-O-)}};
        
        \draw[blue, dashed] (-0.4,-0.4) circle (0.25);
        \draw[blue, dashed] (0.4,-0.4) circle (0.25);
        \draw[blue, dashed] (-0.4,0.4) circle (0.25);
        \draw[blue, dashed] (0.4,0.4) circle (0.25);
        
        \draw[blue, dashed] (0,0) circle (0.816);
    \end{scope}
    

    \begin{scope}[shift={(-3.5,1)}]
        \filldraw[red!70!black, draw=black] (0,0) circle (0.3);
        \fill[red!50!white, opacity=0.6] (0,0) circle (0.2);
        \node at (0,0) {\large\textcolor{white}{$\boldsymbol +$}};
        
        
        
        \draw[blue, dashed, draw opacity=0.4] (0,0) circle (0.4);
    \end{scope}
    
    \begin{scope}[shift={(-3.5,0.3)}]
        \filldraw[red!70!black, draw=black] (0,0) circle (0.3);
        \fill[red!50!white, opacity=0.6] (0,0) circle (0.2);
        \node at (0,0) {\large\textcolor{white}{$\boldsymbol +$}};
        
        
        
        \draw[blue, dashed, draw opacity=0.4] (0,0) circle (0.4);
    \end{scope}
    
    \begin{scope}[shift={(-3.5,-0.4)}]
        \filldraw[red!70!black, draw=black] (0,0) circle (0.3);
        \fill[red!50!white, opacity=0.6] (0,0) circle (0.2);
        \node at (0,0) {\large\textcolor{white}{$\boldsymbol +$}};
        
        
        
        \draw[blue, dashed, draw opacity=0.4] (0,0) circle (0.4);
    \end{scope}
\end{tikzpicture}

%% file: tikzfileGraphicalAbstract.tex

\fontfamily{lmss}

\begin{tikzpicture}[scale=1.5]
    \draw[line width=1pt] (-5,1.25) rectangle (5,-1.25);
    
    \node[font = \large] at (0,-1.41) { distance from electrode / nm};
    \node at (-5,0) [above, rotate=90] {\large\bfseries electrode};
    
    \draw[blue!50!red, line width=2.5pt, domain=0:5, samples=100] plot (\x, {(2/(2+exp(2.5*\x-4))-2)*5/8});
    \draw[blue!50!red, line width=2.5pt, domain=-5:0, samples=100] plot (\x, {(6/(2+exp(2.7*\x+7))-2+2/(2+exp(-4))-6/(2+exp(7)))*5/8}) node[below,font=\Large] at (-2,-0.48) {\bfseries charge density};
    
    \draw[yellow!30!orange, line width=2.5pt, domain=-5:5, samples=100] plot (\x, {5/(2+exp(-2.7*\x-7))-1.25}) node[right, font=\Large] at (-3.6,-1.05) {\bfseries ion-solvent coordination number};

    \begin{scope}[shift={(2.5,0.4)}]
        \filldraw[blue!50!black, draw=black] (0,0) circle (0.3);
        \fill[blue!30!white, opacity=0.6] (0,0) circle (0.2);
        \node at (0,0) {\large\textcolor{white}{$\boldsymbol -$}};
        
        \node at (-0.4,-0.4) {\setchemfig{atom style={scale=0.6}, atom sep=0.35, bond offset = 0.35}\chemfig{[:155]*5(--O-(=O)-O-)}};
        \node at (0.4,-0.4) {\setchemfig{atom style={scale=0.6}, atom sep=0.35, bond offset = 0.35}\chemfig{[:-115]*5(--O-(=O)-O-)}};
        \node at (-0.4,0.4) {\setchemfig{atom style={scale=0.6}, atom sep=0.35, bond offset = 0.35}\chemfig{[:65]*5(--O-(=O)-O-)}};
        \node at (0.4,0.4) {\setchemfig{atom style={scale=0.6}, atom sep=0.35, bond offset = 0.35}\chemfig{[:-25]*5(--O-(=O)-O-)}};
        
        \draw[blue, dashed] (-0.4,-0.4) circle (0.25);
        \draw[blue, dashed] (0.4,-0.4) circle (0.25);
        \draw[blue, dashed] (-0.4,0.4) circle (0.25);
        \draw[blue, dashed] (0.4,0.4) circle (0.25);
        
        \draw[blue, dashed] (0,0) circle (0.816);
    \end{scope}
    
    \begin{scope}[shift={(4.1,-0.4)}]
        \filldraw[red!70!black, draw=black] (0,0) circle (0.3);
        \fill[red!50!white, opacity=0.6] (0,0) circle (0.2);
        \node at (0,0) {\large\textcolor{white}{$\boldsymbol +$}};
        
        \node at (-0.4,-0.4) {\setchemfig{atom style={scale=0.6}, atom sep=0.35, bond offset = 0.35}\chemfig{[:-25]*5(--O-(=O)-O-)}};
        \node at (0.4,-0.4) {\setchemfig{atom style={scale=0.6}, atom sep=0.35, bond offset = 0.35}\chemfig{[:65]*5(--O-(=O)-O-)}};
        \node at (-0.4,0.4) {\setchemfig{atom style={scale=0.6}, atom sep=0.35, bond offset = 0.35}\chemfig{[:-115]*5(--O-(=O)-O-)}};
        \node at (0.4,0.4) {\setchemfig{atom style={scale=0.6}, atom sep=0.35, bond offset = 0.35}\chemfig{[:155]*5(--O-(=O)-O-)}};
        
        \draw[blue, dashed] (-0.4,-0.4) circle (0.25);
        \draw[blue, dashed] (0.4,-0.4) circle (0.25);
        \draw[blue, dashed] (-0.4,0.4) circle (0.25);
        \draw[blue, dashed] (0.4,0.4) circle (0.25);
        
        \draw[blue, dashed] (0,0) circle (0.816);
        
        \node at (-0.2,1.1) {\setchemfig{atom style={scale=0.6}, atom sep=0.35, bond offset = 0.35}\chemfig{[:-70]*5(--O-(=O)-O-)}};
        \node at (-1.4,-0.3) {\setchemfig{atom style={scale=0.6}, atom sep=0.35, bond offset = 0.35}\chemfig{[:140]*5(--O-(=O)-O-)}};
        
    \end{scope}
    
    \begin{scope}[shift={(-1.03,0.35)}]
        \filldraw[red!70!black, draw=black] (0,0) circle (0.3);
        \fill[red!50!white, opacity=0.6] (0,0) circle (0.2);
        \node at (0,0) {\large\textcolor{white}{$\boldsymbol +$}};
        
        \node at (-0.4,-0.4) {\setchemfig{atom style={scale=0.6}, atom sep=0.35, bond offset = 0.35}\chemfig{[:-25]*5(--O-(=O)-O-)}};
        \node at (0.4,-0.4) {\setchemfig{atom style={scale=0.6}, atom sep=0.35, bond offset = 0.35}\chemfig{[:65]*5(--O-(=O)-O-)}};
        \node at (-0.4,0.4) {\setchemfig{atom style={scale=0.6}, atom sep=0.35, bond offset = 0.35}\chemfig{[:-115]*5(--O-(=O)-O-)}};
        \node at (0.4,0.4) {\setchemfig{atom style={scale=0.6}, atom sep=0.35, bond offset = 0.35}\chemfig{[:155]*5(--O-(=O)-O-)}};
        
        \draw[blue, dashed] (-0.4,-0.4) circle (0.25);
        \draw[blue, dashed] (0.4,-0.4) circle (0.25);
        \draw[blue, dashed] (-0.4,0.4) circle (0.25);
        \draw[blue, dashed] (0.4,0.4) circle (0.25);
        
        \draw[blue, dashed] (0,0) circle (0.816);
    \end{scope}
    
    \begin{scope}[shift={(0.6,0.2)}]
        \filldraw[red!70!black, draw=black] (0,0) circle (0.3);
        \fill[red!50!white, opacity=0.6] (0,0) circle (0.2);
        \node at (0,0) {\large\textcolor{white}{$\boldsymbol +$}};
        
        \node at (-0.4,-0.4) {\setchemfig{atom style={scale=0.6}, atom sep=0.35, bond offset = 0.35}\chemfig{[:-25]*5(--O-(=O)-O-)}};
        \node at (0.4,-0.4) {\setchemfig{atom style={scale=0.6}, atom sep=0.35, bond offset = 0.35}\chemfig{[:65]*5(--O-(=O)-O-)}};
        \node at (-0.4,0.4) {\setchemfig{atom style={scale=0.6}, atom sep=0.35, bond offset = 0.35}\chemfig{[:-115]*5(--O-(=O)-O-)}};
        \node at (0.4,0.4) {\setchemfig{atom style={scale=0.6}, atom sep=0.35, bond offset = 0.35}\chemfig{[:155]*5(--O-(=O)-O-)}};
        
        \draw[blue, dashed] (-0.4,-0.4) circle (0.25);
        \draw[blue, dashed] (0.4,-0.4) circle (0.25);
        \draw[blue, dashed] (-0.4,0.4) circle (0.25);
        \draw[blue, dashed] (0.4,0.4) circle (0.25);
        
        \draw[blue, dashed] (0,0) circle (0.816);
    \end{scope}
    
    
    \begin{scope}[shift={(-3.7,0.3)}]
        \filldraw[red!70!black, draw=black] (0,0) circle (0.3);
        \fill[red!50!white, opacity=0.6] (0,0) circle (0.2);
        \node at (0,0) {\large\textcolor{white}{$\boldsymbol +$}};
        
        \node at (0.4,-0.4) {\setchemfig{atom style={scale=0.6}, atom sep=0.35, bond offset = 0.35}\chemfig{[:65]*5(--O-(=O)-O-)}};
        \node at (0.4,0.4) {\setchemfig{atom style={scale=0.6}, atom sep=0.35, bond offset = 0.35}\chemfig{[:155]*5(--O-(=O)-O-)}};
        
        \draw[blue, dashed] (-0.4,-0.4) circle (0.25);
        \draw[blue, dashed] (0.4,-0.4) circle (0.25);
        \draw[blue, dashed] (-0.4,0.4) circle (0.25);
        \draw[blue, dashed] (0.4,0.4) circle (0.25);
        
        \draw[blue, dashed] (0,0) circle (0.816);
    \end{scope}
    

    \begin{scope}[shift={(-4.55,0.7)}]
        \filldraw[red!70!black, draw=black] (0,0) circle (0.3);
        \fill[red!50!white, opacity=0.6] (0,0) circle (0.2);
        \node at (0,0) {\large\textcolor{white}{$\boldsymbol +$}};
        
        
        
        \draw[blue, dashed, draw opacity=0.4] (0,0) circle (0.4);
    \end{scope}
    
    \begin{scope}[shift={(-4.55,0)}]
        \filldraw[red!70!black, draw=black] (0,0) circle (0.3);
        \fill[red!50!white, opacity=0.6] (0,0) circle (0.2);
        \node at (0,0) {\large\textcolor{white}{$\boldsymbol +$}};
        
        
        
        \draw[blue, dashed, draw opacity=0.4] (0,0) circle (0.4);
    \end{scope}
    
    \begin{scope}[shift={(-4.55,-0.7)}]
        \filldraw[red!70!black, draw=black] (0,0) circle (0.3);
        \fill[red!50!white, opacity=0.6] (0,0) circle (0.2);
        \node at (0,0) {\large\textcolor{white}{$\boldsymbol +$}};
        
        
        
        \draw[blue, dashed, draw opacity=0.4] (0,0) circle (0.4);
    \end{scope}
    
\end{tikzpicture}
